# ON STABILITY OF PHYSICAL SYSTEMS IN A COMPLEX PLANE

V.V. Lyahov, V.M. Neshchadim

## 1. Introduction

Complex numbers have fully come into common use in mathematics from the very moment of their discovery. Let's mention only one their property – the property of algebraic closedness of the field of complex numbers. The physics even now is in the stage of half-recognizing of complex numbers. On the one hand, the physics, using mathematical means, manipulates imaginaries; on the other hand, common sense says that all observable quantities should be associated with real numbers. Complex calculus should be only an auxiliary, mathematically formal tool. The line of reasoning of a modern physicist is as follows: “We live in the real world; therefore, all quantities should be featured by real numbers”. This statement seems to be completely natural without any additional argumentation. Such situation is an example of paradigm and is taken for granted by scientists. It may be said that the advancement of science confirms this thesis.

But it seems that there are also opposite examples.

Imaginary time$\tau = i \cdot ct$ is introduced into the relativity theory, and only in this form together with three spatial coordinates it forms a four-dimensional space-time. The fourth axis of space-time is an imaginary quantity. In the theory of relativity, on the one hand, it is always underlined that imaginary time is of conditional nature, but, on the other hand, it is also always noted that a deeply essential linkage between space and time was discovered for the first time.

The basic object of the quantum mechanics - a complex-valued wave function is referred to as not being of any physical sense, but the square of this function has physical sense. This conclusion alone causes concern; moreover, it appears that this wave function and not its square with physical sense should appear in the quantum mechanical principle of superposition.

Because of mathematical reasons about necessity of completeness of a system of wave functions, a notion of negative energy levels was introduced into the relativistic quantum mechanics. But energy of a resting particle $E=mc^2$ may be negative only in the event that a negative value is assigned to mass, or an imaginary value is assigned to velocity of light. . Physical interpretation of this formal mathematical result is given by Dirac who postulated principal unobservability of states with negative energy due to the fact that all levels with negative energy were occupied by particles, and consequently any transitions between any two levels which might be observed, are impossible (state vacuum). But, postulating in the beginning principal unobservability of such states, physicists speak then about interaction of atom of hydrogen with vacuum (Lamb shift). It means that the background of filled states with negative energy- vacuum do manifests itself in reality, isn't it?

And finally, a long known thing. A physicist when solving an algebraic equation with real coefficients, gains frequently complex roots. How can we treat them? Solutions are selected according to a so-called physical sense. A rigorous mathematical solution is exposed to violence by virtue of any prior knowledge. Paths of physics and mathematics diverge. Should they diverge? Is physical sense infallible, being a sense only now and here and giving up the place to another sense eventually?

Probably it's time for physics to introduce complex numbers into common use, i.e. to make all the observed quantities correspond to complex numbers instead of real numbers.

## 2. Formulation of the problem

Immediate use of the non-Archimedian complex numerical axes entered into consideration [1, 2] is still impossible. The technique of differentiation and integration on these objects is to be developed. Besides, it is not known, what laws act through inenumerable infinite distances or times of a new «complex-valued reality»? Do physical laws known to us now operate there, or are they somehow transformed? Therefore, only real segments $\alpha_o$ of the central part of the non-Archimedian complex numerical axes of space and time [2] interpreted as a usual «real world» with all known physical laws will be considered in the further examination. We guess that in this actual space-time the complex-valued nature of the remaining physical quantities may be manifested.

All the above, in our opinion, necessitates complexification of physics, and only further theoretical consideration will show the consequences. The complexification properly lies in the assumption that quantities treated in all physical laws, are complex, and their imaginary parts are small. Theoretically we shall try to determine whether the situation when solution for the real part of the entered complex quantities essentially differs from the ordinary true solution defined by the equations of any physical law is possible.

The subject of our examination will be the differential equations, which are the basis of modeling of physical processes. Solution of any differential equation can be built either on a real axis or in a complex plane. In this case, the first solution is a special case of the second more general solution since a real number may be derived from the complex one when the imaginary part of the latter is made vanish. In practice, not a single quantity may be precisely measured or set, since all quantities fluctuate. By now in the field of real numbers the stability theory based on Lyapunov approach [3] and permitting to explore the dependence of solution of a problem on fluctuation of additional conditions is developed.

The paper offers to apply the technique of examination of the Lyapunov stability to the complex plane to study the dependence of the solution of the problem on the fluctuation of imaginary parts of additional conditions in the vicinity of zero. Let's solve the Cauchy problem analyzing how the complex solution transforms into a real one, with imaginary parts approaching zero. Will the continuity of dependence of the solution on change of the starting conditions remain in this case? It's reasonable to begin with the most simple objects and laws of physics solving subproblems.

## 3. Examination of equations of the classical physics

Any equation contains three kinds of elements: functions, arguments, and coefficients. Formulation of the problem necessitates in general considering all the elements as complex-valued, but solving the nonlinear complex-valued equations we shall take into consideration only functions, since more simple linear equations makes it also possible to consider the influence of complex-valued quantities in arguments and coefficients.

a) Linear first-order equation with constant coefficients

The equation

$$\frac{dN}{dt} = -\lambda \cdot N \tag{3.1}$$

features phenomenon of radioactivity, light absorption, gamma-ray absorption, equilibriums of phases (Clapeyron-Clausius equation). This is an autonomous equation; $N$ function and $\lambda$ parameter are complex-valued.

Let's consider an initial value problem (Cauchy problem)

$$N(t=0) = N_0 \tag{3.2}$$

with the purpose of examination of the influence of small imaginary parts of initial conditions and parameters on the real part of the solution.

The solution of the problem (3.1), (3.2)

$$N = N_0 \exp(-\lambda \cdot t) \tag{3.3}$$

is examined numerically in the complex domain. The results of the solution are shown in Fig. 1÷5. Calculations are carried out for half-life period $T_{1/2}= 10$ days ($\lambda= ln2/T_{1/2}$). Relations of imaginary parts to actual parts for all quantities were taken identical:

$$\frac{\operatorname{Im} N_0}{\operatorname{Re} N_0} = \frac{\operatorname{Im} \lambda}{\operatorname{Re} \lambda} = \frac{J}{R} \;;$$

Figure 1a (*J/R= 0*) demonstrates the ordinary radioactive decay law. The phase-plane portrait in {*N*, *dN/dt*} coordinates shown in Fig. 1b represents a stable node.

Figures 2a and 2b, (*J/R= 0.3*), explicitly reveals deviation from the known law of radioactive decay. As *J/R* parameter crosses zero, the first bifurcation occurs and a stable node transforms into a stable focus.

Figure 3, (*J/R= 0.9*), represents a well expressed stable focus.

At the *J/R= 1* value (refer to Figure 4) the second bifurcation happens - a Hopf bifurcation when a stable focus turns into a limit cycle.

At the *J/R= 1.01* value (refer to Figure 5) the third bifurcation takes place when the limit cycle turns into an unstable focus.

As far as we know, a solution bifurcation phenomenon was always associated with nonlinearity of equations, but the most elementary linear equation (3.1) of the possible ones in a complex plane is also seen to exhibit a bifurcation property.

The above second and third bifurcations are observed at large *J/R* values and, therefore, are of mathematical interest only. But the first bifurcation that reveals itself only in the vicinity of *J/R=0* may have a physical content. Even small imaginary parts of coefficients and starting conditions ($J/R \approx 0$) result in qualitative reorganization of the real part of the solution transforming a stable node into a stable focus.

b) Linear second-order equation with constant coefficients

The equation

$$\frac{d^2 y}{dt^2} + \frac{\gamma}{m}\frac{dy}{dt} + \frac{\kappa}{m} = 0 \qquad (3.4)$$

features a mechanical oscillator and an electrical oscillating circuit. A function of $y$ - bias of a particle and parameters: $m$ - a particle mass, $\gamma$ - a friction coefficient, $\kappa$ - an elasticity coefficient are complex-valued quantities. As starting conditions we shall take:

$$\begin{cases} y(t = 0) = \delta y \\ \dot{y}(t = 0) = 0 \end{cases} \qquad (3.5)$$

The solution of the Cauchy problem (3.4), (3.5) is known:

$$y = c_1 \cdot e^{k_1 t} + c_2 \cdot e^{k_2 t} \tag{3.6}$$

Here, $k_{1,2}$ are roots of a characteristic equation:

$$k_{1,2} = -\frac{\gamma}{2m} \pm \sqrt{\frac{\gamma^2}{4m^2} - \frac{\kappa}{m}}, \tag{3.7}$$

Coefficients $c_1$ and $c_2$ are determined from the starting conditions (3.5):

$$c_1 = \frac{\delta y}{1 - \frac{k_1}{k_2}}; \qquad c_2 = -\frac{k_1}{k_2} \frac{\delta y}{1 - \frac{k_1}{k_2}}. \tag{3.8}$$

As shown from the formula (3.7) three parameters $m$, $\kappa$, $\gamma$ are used for defining the roots of the characteristic equation. Numerical examination of solutions (3.6) in the complex domain shows that beginning from a certain quantity of ratios

$$\frac{\operatorname{Im} N_0}{\operatorname{Re} N_0} = \frac{\operatorname{Im}\kappa}{\operatorname{Re}\kappa} = \frac{\operatorname{Im}\gamma}{\operatorname{Re}\gamma}$$

and if they continue to increase, a root with a positive real part appears among roots $k_{1,2}$, which indicates that the system (3.4), (3.5) lost its stability. In the real domain the solution of the problem (3.4), (3.5) is always stable. It may be said that solution of the problem (3.4), (3.5) in six-measured space of parameters {Re$m$, Im$m$, Re$\kappa$, Im$\kappa$, Re$\gamma$, Im$\gamma$} is stable everywhere, except for some domains. Only those domains of instability are of physical interest where imaginary quantities are small:

$$\frac{\operatorname{Im} N_0}{\operatorname{Re} N_0} << 1, \qquad \frac{\operatorname{Im}\kappa}{\operatorname{Re}\kappa} << 1, \qquad \frac{\operatorname{Im}\gamma}{\operatorname{Re}\gamma} << 1 \quad .$$

The analysis of expressions (3.7) shows that the solution ceases to be stable, that is a real part of one of the roots (3.7) becomes positive (particularly equal $0.5 \cdot 10^{-10}$ HZ) at

$$\begin{cases} \dfrac{\operatorname{Im} m}{\operatorname{Re} m} = \dfrac{\operatorname{Im}\kappa}{\operatorname{Re}\kappa} = \dfrac{\operatorname{Im}\gamma}{\operatorname{Im}\gamma} \geq 10^{-6}, \\ if \quad \operatorname{Re} m = 10g; \quad \operatorname{Re}\kappa = 1000 Hz^2 \cdot g; \quad \operatorname{Re}\gamma = 0.005 Hz \cdot g \end{cases} \tag{3.9}$$

and becomes equal to *0.005 Hz* already at

$$\begin{cases} \dfrac{\operatorname{Im} m}{\operatorname{Re} m} = \dfrac{\operatorname{Im}\kappa}{\operatorname{Re}\kappa} = \dfrac{\operatorname{Im}\gamma}{\operatorname{Im}\gamma} \geq 10^{-12}, \\ if \quad \operatorname{Re} m = 10g; \quad \operatorname{Re}\kappa = 10^{21} \cdot Hz^2 \cdot g; \quad \operatorname{Re}\gamma = 0.05 \cdot Hz \cdot g \end{cases} \tag{3.10}$$

Solution for mechanical pendulum also has the feature that the continuity of dependence of the solution on change of an imaginary part of parameters in the neighborhood of zero is broken; the solution through a bifurcation passes from a stable type to a qualitative another unstable type as soon as the imaginary part of parameters exceeds a certain value. A phase-plane portrait of the solution at (3.10) parameters and $J/R < 10^{-12}$ value is shown in Figure 6a and represents a slowly convergent focus. The phase-plane portrait for the same parameters (3.10) and $J/R > 10^{-12}$ values represent a slowly divergent focus (refer to Figure 6b). Focuses very slowly converge and diverge because a friction coefficient $\gamma$ and, as a consequence, the real parts of characteristic roots (3.7) are very small.

c) Other equations of physics (a rotator, the Newton equation with constant force, the Kepler problem)

The flat rotator is described by the equation

$$\frac{d^2\varphi}{dt^2} = 0, \tag{3.11}$$

Its solution is

$$\varphi = c \cdot t + \varphi_0. \tag{3.12}$$

If quantities c, $\varphi_0$ are complex, the complex solution can be written in the explicit form:

$$\varphi = (\mathrm{Re}c \cdot t + \mathrm{Re}\varphi_o) + i(\mathrm{Im}c \cdot t + \mathrm{Im}\varphi_o). \tag{3.13}$$

It can be seen that the real part of the complex solution (3.13) differs from the strongly real solution (3.12) only quantitatively, but belongs to the same type. When imaginary parts of quantities approach zero $\mathrm{Im}c \to 0$, $\mathrm{Im}\varphi_o \to 0$, the real part of the solution (3.13) continuously passes into the real solution (3.12).

The known solution of the Newton equation with constant force

$$m \cdot \frac{d^2x}{dt^2} = F \tag{3.14}$$

looks like

$$x = \frac{F}{2m} \cdot t^2 + c_1 \cdot t + c_2. \tag{3.15}$$

If quantities $F$, $m$, $c_1$, $c_2$ are complex, then through introduction of an explicit denotation

$$\frac{F}{2m} = A + iB, \qquad c_1 = \mathrm{Re}\, c_1 + i\,\mathrm{Im}\, c_1, \qquad c_2 = \mathrm{Re}\, c_2 + i\,\mathrm{Im}\, c_2,$$

the solution (3.15) can be rewritten as:

$$x = A \cdot t^2 + \mathrm{Re}\, c_1 \cdot t + \mathrm{Re}\, c_2 + i[B \cdot t^2 + \mathrm{Im}\, c_1 \cdot t + \mathrm{Im}\, c_2] \quad (3.16)$$

As in the case of the rotator, the real part of the complex solution (3.16) differs from the real solution (3.15) only quantitatively, but belongs to the same type, that is the solution is stable with respect to complex-valued perturbations.

The Kepler problem represents a problem of two bodies, between which the force of interaction is inversely proportional to the squared distance; as this force may be the gravitational force or Coulomb force. The Kepler problem is represented by the equation of Newton which right member is inversely proportional to the squared distance. In polar coordinates the solution of the problem has the following form:

$$r = \frac{P}{1 + e \cdot \cos\varphi} \quad (3.17)$$

Parameter of an orbit $P$ and eccentricity of an orbit $e$ in case of gravitation, for example, are expressed through parameters of the problem as follows:

$$P = \frac{L^2}{\gamma \cdot m^2 \cdot M}, \qquad e = \sqrt{1 + \frac{2 \cdot W \cdot L^2}{\gamma^2 \cdot m^3 \cdot M^2}};$$

Here, $\gamma$ - a gravitational constant, $m$ - mass of a light body, $M$ - mass of a heavy body, $L$ and $W$ - a moment of momentum and a total energy of a light body, respectively.

We assume that function $r$, argument $\varphi$ and parameters $P$ and $e$ are complex-valued. The solution (3.17) was analyzed numerically to define the influence of complex-valued perturbations of parameters $P$ and $e$ on the real part of the solution.

It is known that in the real domain the solution (3.17) represents a hyperbola if $e > 1$, a parabola if $e = 1$, and an ellipse if $e < 1$. The fluctuation of parameter $e$ in the neighborhood of unity may throw an orbit from one kind of a trajectory on another. Our concern is only with the fact whether changes of the real solution caused by fluctuation of the real parameter in the neighborhood of unity differ from those of the real part of the solution which may be caused by fluctuation of the imaginary part of parameter *Ime* when *Ree = 1*.

Solutions (3.17) in the real domain are shown in Figure 7÷10. Calculations were made for parameter $P = 1$. Trajectories are shown in coordinates $x = r \cdot \cos\varphi$, $y = r \cdot \sin\varphi$. It may be seen that change of the parameter $e$ from *0.9* to *1.1* results in change of the trajectory from ellipse to hyperbola remaining visually still an ellipse at $e_{кр} = 0.(9)^4$. Change of parameter $e_{кр} = 0.(9)^4$ to one or another side makes a trajectory either an ellipse or a parabola.

Figures 11 and 12 show change of the real part of the solution (3.17) when *Ree*= *0.(9)*$^4$ and increase of perturbation of the imaginary part *J/R* from $10^{-4}$ to *0.01*. We assume that

$$\frac{J}{R}=\frac{\operatorname{Im}P}{\operatorname{Re}P}=\frac{\operatorname{Im}e}{\operatorname{Re}e}=\frac{\operatorname{Im}\varphi}{\operatorname{Re}\varphi}.$$

It is seen that the critical ellipse is deformed not into a parabola as it was in case of fluctuations of the real value *e* but into a completely another curve.

Solution of the problem (3.17) in the neighborhood of *Ree = 1* is unstable both as to real perturbations *Ree* and as to imaginary perturbations *Ime*, but fluctuations *Ree* and *Ime* yield completely different results with different trajectory curves.

d) Nonlinear equations of the plasma physics

Let's study the solution of two Cauchy problems in the complex domain of variation. The first problem represents a system of equations:

$$\begin{cases}\lambda\dfrac{d^2\psi}{d\xi^2}=\exp(\psi-\dfrac{\alpha_e}{1+\alpha_e}\cdot\dfrac{a^2}{2\mu})-\exp(-\psi-\dfrac{\alpha_i}{1+\alpha_i}\cdot\dfrac{a^2}{2}); \\ \dfrac{d^2a}{d\xi^2}=\dfrac{\alpha_e}{1+\alpha_e}\cdot\dfrac{a}{\mu}\cdot\exp(\psi-\dfrac{\alpha_e}{1+\alpha_e}\cdot\dfrac{a^2}{2\mu})+\dfrac{\alpha_i\cdot a}{1+\alpha_i}\cdot\exp(-\psi-\dfrac{\alpha_i}{1+\alpha_i}\cdot\dfrac{a^2}{2})\end{cases} \quad (3.18)$$

with boundary conditions as follows:

$$\begin{cases}\operatorname{Re}\psi(0)=c_1,\quad \operatorname{Im}\psi(0)=c_2,\quad \dfrac{d\operatorname{Re}\psi(0)}{d\xi}=c_3,\quad \dfrac{d\operatorname{Im}\psi(0)}{d\xi}=c_4, \\ \operatorname{Re}a(0)=c_5,\quad \operatorname{Im}a(0)=c_6,\quad \dfrac{d\operatorname{Re}a(0)}{d\xi}=c_7,\quad \dfrac{d\operatorname{Im}a(0)}{d\xi}=c_8.\end{cases} \quad (3.19)$$

The second Cauchy problem consists of the equations:

$$\begin{cases}\dfrac{d^2\psi}{d\xi^2}=\dfrac{1}{[1+2(1-\gamma)\cdot\psi-(1-\gamma)^2\cdot a^2]^{1/2}}-\dfrac{1}{[1-2\gamma\cdot\psi-\gamma^2\cdot a^2]^{1/2}}; \\ \dfrac{d^2a}{d\psi^2}=\eta^2\left\{\dfrac{(1-\gamma)\cdot a}{[1+2(1-\gamma)\psi-(1-\gamma)^2a^2]^{1/2}}+\dfrac{\gamma\cdot a}{[1-2\gamma\psi-\gamma^2a^2]^{1/2}}\right\},\end{cases} \quad (3.20)$$

with boundary conditions:

$$\begin{cases} \operatorname{Re}\psi(\xi_0)=c_1, \quad \operatorname{Im}\psi(\xi_0)=c_2, \quad \operatorname{Re}a(\xi_0)=c_3, \quad \operatorname{Im}a(\xi_0)=c_4, \\ \dfrac{d\operatorname{Re}\psi(\xi_0)}{d\xi}=c_5, \quad \dfrac{\operatorname{Im}\psi(\xi_0)}{d\xi}=c_6, \quad \dfrac{d\operatorname{Re}a(\xi_0)}{d\xi}=c_7, \quad \dfrac{d\operatorname{Im}a(\xi_0)}{d\xi}=c_8. \end{cases} \tag{3.21}$$

The equations (3.18) and (3.20) were derived in papers [4] and [5] and represent a system of Maxwell equations for potentials of a self-consistent electromagnetic field in stationary active magnetic Maxwell and cold plasma, respectively. The equations are written in the dimensionless form, $\psi$ and $a$ - electrical and magnetic potentials, $\alpha_{e,i}= (T_{1e,i}/T_{\Pi e,i}) - 1$ - a degree of plasma anisotropy, where $T_{1e,i}$ - temperature of the plasma relevant component across a polarization magnetic field and along a polarization electric field, $T_{\Pi e,i}$ - temperature across electrical and magnetic fields $\lambda= T_{1i}/m_ic^2$, $\mu= m_e/m_i$, $\gamma= m_e/(m_e+m_i)$, $\eta= v_o/c$, in which $m_e$ and $m_i$ - masses of an electron and an ion, $v_o$ - velocity of a stream of cold plasma, $c$ - velocity of light.

The equations (3.18) and (3.20) are systems of nonlinear autonomous differential equations. Let’s first examine their equilibrium and stability on a real axis.

Let's present the equations (3.18) as:

$$\begin{cases} \dfrac{d\psi}{d\xi}= & \psi'; \\ \dfrac{d\psi'}{d\xi}= & \dfrac{1}{\lambda}[\exp(\psi-\dfrac{\alpha_e}{1+\alpha_e}\cdot\dfrac{a^2}{2\mu})-\exp(-\psi-\dfrac{\alpha_i}{1+\alpha_i}\cdot\dfrac{a^2}{2})]; \\ \dfrac{da}{d\xi}= & a'; \\ \dfrac{da'}{d\xi}= & a[\dfrac{1}{\mu}\cdot\dfrac{\alpha_e}{1+\alpha_e}\cdot\exp(\psi-\dfrac{\alpha_e}{1+\alpha_e}\cdot\dfrac{a^2}{2\mu})+\dfrac{\alpha_i}{1+\alpha_i}\cdot\exp(-\psi-\dfrac{\alpha_i}{1+\alpha_i}\cdot\dfrac{a^2}{2})] \end{cases} \tag{3.22}$$

Equality of the right members to zero determines equilibrium points:

$$\psi=\psi'=a=a'. \tag{3.23}$$

Linearized system in the neighborhood of equilibrium [6]

$$\frac{\partial y_i}{\partial t}=\sum_{k=1}^{4}\left(\frac{\partial f_i}{\partial y_k}\right)_0(y_k-y_{ko}), \quad i=1,2,3.4$$

for the equations (3.22) in the neighborhood of an equilibrium point (3.23) is as follows:

$$\begin{cases} \dfrac{d\psi}{d\xi} = & \psi'; \\ \dfrac{d\psi'}{d\xi} = & \dfrac{2}{\lambda}\cdot\psi; \\ \dfrac{da}{d\xi} = & a'; \\ \dfrac{da'}{d\xi} = & \left(\dfrac{1}{\mu}\cdot\dfrac{\alpha_e}{1+\alpha_e}+\dfrac{\alpha_i}{1+\alpha_i}\right)\cdot a, \end{cases} \tag{3.24}$$

The characteristic equation for the first pair of equations (3.24) looks like**:**

$$\begin{vmatrix} -r & 1 \\ \dfrac{2}{\lambda} & -r \end{vmatrix} = 0,$$

its roots are:

$$r_{1,2} = \pm\sqrt{\frac{2}{\lambda}} \tag{3.25}$$

It means that solution of the first pair of the equations (3.24) in the neighborhood of a stationary point is of a saddle type.

Solution of the characteristic equation for the second pair of equations of the system (3.24) also determines a saddle:

$$r_{3,4} = \pm\sqrt{\frac{1}{\mu}\cdot\frac{\alpha_e}{1+\alpha_e}+\frac{\alpha_i}{1+\alpha_i}}. \tag{3.26}$$

Among roots $r_{1;2;3;4}$ there are roots with a positive real part, i.e. the equilibrium point (3.23) is unstable.

Equilibrium and stability of the system (3.20) are similarly examined. It appears that the equilibrium point of this system is also determined by expression (3.23), and the roots of the characteristic equation determine by themselves a center:

$$r_{5,6} = \pm i\sqrt{\frac{1}{\mu}\cdot\frac{\alpha_e}{1+\alpha_e}+\frac{\alpha_i}{1+\alpha_i}}. \tag{3.27}$$

and a saddle:

$$r_{7,8} = \pm\sqrt{1+\mu} \tag{3.28}$$

Equilibrium of the system (3.20) is also unstable.

Let's now pass to solution of the problems (3.18), (3.19), (3.20), and (3.21).

Only the sought functions will be considered as complex-valued.

The problems (3.18), (3.19), (3.20), and (3.21) are too complicated to be solved analytically; therefore, properties of their solutions were examined numerically using the Gear method for integration of stiff systems of equations. This method provided a self-selection of an integration step and calculation with prescribed accuracy [7-11].

Let's consider results of the solution of the problem (3.18), (3.19). If $c_2 = c_4 = c_6 = c_8 = 0$ is taken in the boundary conditions (3.19), then the remaining constants $c_1$, $c_3$, $c_5$, $c_7$ will determine the real solution. Passage to the limit from complex solutions to the real ones is examined only in terms of the constant $\mathrm{Im}\psi(0)= c_2$ by its change from 1 to 0. Calculations carried out in a wide range of parameters $\lambda$, $\alpha$ have shown that, when $|\mathrm{Im}\psi(0)| < 10^{-2}$, the solution always has exponential, hyperbolic type which real part differs little from the strict real solution (refer to Figure 13), and when $|\mathrm{Im}\psi(0)| \geq 10^{-2}$, the solution has entirely different sinusoidal form (refer to Figure14). Since an exponential curve cannot continuously pass into a sine curve, between the two types of solution there is a point of bifurcation.

Table 1 is obtained for $\mathrm{Im}\psi(0) = 10^{-3}$ at the following values of parameters and boundary conditions:

$$\mu= 0.00055;\ \lambda= 0.1;\ \alpha_e= \alpha_i= \alpha= 100;\ c_1= c_5= 0.001;$$

$$c_3= c_4= c_6= c_7= c_8= 0.$$

Table 2 is obtained for $\mathrm{Im}\psi(0)= 10^{-2}$ at the same parameters and boundary conditions. Figure 13 shows profile lines of actual parts from Table 1. Figure 14 shows the same from Table 2.

The nonlinear system (3.18), (3.19) exhibits a strong dependence on parameters, and, as it was shown in numerical experiment, there are such their values $\lambda= 0.1$, $\alpha= 0.01$ which shift a boundary point which is a point of bifurcation between two types of solutions, sinusoidal and exponential, to zero by five orders from value $\mathrm{Im}\psi(0)= 10^{-2}$ to value $\mathrm{Im}\psi(0)= 10^{-7}$. When boundary conditions have these values, even small fluctuations of the imaginary part $\mathrm{Im}\psi(0)$ in the neighborhood of zero may transfer the solution of the system (3.18), (3.19) from the exponential domain to a sinusoidal domain which is qualitatively different.

The conducted numerical examination shows that transition from the complex domain of solutions to the real domain, when $\mathrm{Im}\psi(0) \rightarrow 0$, is not trivial. The continuity of solution dependence on change of the imaginary part of a boundary condition in the neighborhood of zero is violated.

Let's consider the problem (3.20), (3.21). It should be noted that the desire to solve the equation (3.20) in the domain of real numbers requires extension of definition of the right members of the equations: there is such point $\xi = \xi_e$ for which $1 + 2(1 - \gamma)\psi - (1 - \gamma)^2a^2 > 0$ when $\xi < \xi_e$ and for which $1 + 2(1 - \gamma)\psi - (1 - \gamma)^2a^2 < 0$ when $\xi > \xi_e$; the point $\xi = \xi_e$ is considered as a point of reflection of electrons, and in the domain $\xi > \xi_e$, where density of electrons $[1 + 2(1 - \gamma)\psi - (1 - \gamma)^2a^2]^{1/2}$ becomes an imaginary quantity, the latter is replaced by zero in the equations (3.20). Similarly, the point $\xi = \xi_i$, where $1 - 2\gamma\psi - \gamma^2a^2 = 0$, is considered as a point of reflection of ions.

Desire to maintain physically interpreted result when we build the solution of the equations (3.20) in the real domain involves the necessity of extension of their definition. In this case the problem (3.20), (3.21) ceases to obey to the classical theorem of existence and uniqueness of solution. Besides, in the points, defined as points of reflection, density of the relevant components of plasma diverges. And, as it is found out in the paper [5], when $\eta > 0.45$, electrons are first to reflect in the stream of plasma falling on magnetic field, and when $\eta < 0.40$ – ions. This contradicts to intuitive concept that heavier ions in a single-velocity stream should penetrate into the region of the magnetic field more deeply then electrons, and thus, the quantity related to the Debye length should be a characteristic spatial size of charge separation.

Results of the numerical solution of the problem (3.20), (3.21) in the complex plane are shown in Tables 3, 4, 5 and in the relevant Figures 15 and 16. The influence of small imaginary parts of boundary conditions was examined in terms of the component Im$\psi(\xi_o)$. Table 3 shows calculations at Im$\psi(\xi_o) = 0$, Table 4 — at Im$\psi(0) = 10^{-15}$, and Table 5 — at Im$\psi(0) = 10^{-13}$, other values of parameters and boundary conditions are identical:

$$\eta = 0.5;\ \gamma = 0.25;\ c_1 = c_5 = 2.5764\cdot 10^{-10};$$
$$c_3 = 4.5400\cdot 10^{-5};\ c_4 = c_6 = c_7 = c_8 = 0.$$

Comparison of Tables 4 and 5 demonstrates that small change of the imaginary part of boundary conditions results in small change of the solution, and modules of the imaginary parts are always well below than modules of the real parts.

But real parts of the complex solutions shown in Tables 4 and 5 radically differ from the real solution of Table 3. As can be seen, solution of the problem (3.20), (3.21) in a complex plane overcome all difficulties of purely real solution: first, the problem of extension of definition of the right members of input equations disappears; second, density of the components does not become infinite anywhere; third, the situation when ions are the first to be reflected since both ions and electrons simultaneously penetrate into the region of the magnetic field in such a manner that their densities smoothly approach zero does not arise.

Thus, transition of the solution from a complex neighborhood to a real axis has singularity both in the problem (3.18), (3.19) and in the problem (3.20), (3.21). A small imaginary part of an additional condition may lead the problem solution in quite opposite directions depending on whether it more or less than some critical value.

## 4. Summary

The basic and most explicit laws of the classical physics were explored, explicit in that involved quantities are strictly defined with the least degree of subjectivity and may be measured with the high degree of accuracy.

Simple objects, such as a rotator, the Newton equation with constant force, are stable with regard to complex-valued perturbations in the neighborhood of the real axis. As the complexity of objects increase, stability is lost.

Fluctuations of the imaginary part of starting conditions in the neighborhood of zero in the nonlinear equations lead to bifurcations of the real part of the solution. Fluctuations of the imaginary part of coefficients in the neighborhood of zero in the linear equations (the law of radioactive decay, an oscillator) also lead to bifurcations of the real part of the solution.

The conducted examination of nonlinear equations and, which is more important, linear differential equations from the various fields of physics shows that transition of solution from the complex plane to the real axis is not trivial. This transition may be accompanied by violation of the continuous dependence of the solution on the change of imaginary parts of additional conditions and parameters in the neighborhood of zero. In other words, it was revealed that a small imaginary part may control solution on the real axis.

This fact is interesting from standpoint of the formal mathematics, and even more actual is its interpretation in terms of physics and philosophy.

Conceavably, a small imaginary part, even if unobservable, is still an inherent characteristic of a physical quantity, being yet something like a hidden parameter, and manifests itself only indirectly forcing the system to move in this or that direction.

From the point of view of mathematics it is of interest that even simple equations in a complex plane have the property of bifurcation of the real part of the solution. It may be caused by the fact that physical quantities that form the coefficients of equations are complex (vectors), which actually results in nonlinearity of these equations as to the real parts.

## 5. Implementation of an experiment

As a possible experiment to check the hypothesis concerning a complex-valued nature of physical quantities put forward, the following experiment with an oscillator is offered. We may hope to find out this effect only in the fields far from ordinary practice. So, properties (3.10) which are required for the oscillator to reveal the imaginary quantities of the order of

$$\frac{\operatorname{Im} m}{\operatorname{Re} m} = \frac{\operatorname{Im} \kappa}{\operatorname{Re} \kappa} = \frac{\operatorname{Im} \gamma}{\operatorname{Re} \gamma} = \frac{J}{R} \geq 10^{-12}$$

lay on the verge of the current technological capabilities [12]. The oscillator should have the greatest possible natural frequency $\omega \approx (\kappa/m)^{1/2} = 10^{10}$ *Hz* and minimal dissipation $\gamma = 0.05$ *Hz·g*, which requires cryogenic and vacuum technique. Analysing a Cauchy problem for such oscillator, one may try to find out the existence of imaginary quantities (their influence on the real part of the solution) if they are not less than $J/R = 10^{-12}$, which is physically rather small value.

In a series of experiences that demonstrate an ordinary convergent solution at $J/R < 10^{-12}$ (refer to Figure 6a) it may be revealed a certain number of divergent solutions (refer to Figure 6b) if fluctuations of parameters reach value $J/R > 10^{-12}$ and if, certainly, physical quantities are complex-valued. Even if oscillations diverge, it may manifest itself only within a certain initial time period since at a later time a test sample will heat up as a result of oscillations, the dissipation will increase, and the system will transform into an ordinary mode of convergent oscillations.

Revealing the complexity of physical quantities in principle provides for experiment within the range of a linear equation of the first order (the low of radioactive decay, low of light and gamma-ray absorption, the equation of phase equilibrium) and experimental investigation of stability of the Kepler problem as applied to the Coulomb interaction.

Within the framework of the hypothesis offered by authors about a complex-valued nature of physical quantities the stability of basic equations of the classical physics concerning complex-valued perturbations of parameters and boundary conditions is explored. The conducted examination of nonlinear equations and, what is more important, of linear differential equations shows violation in some cases of the continuous dependence of the solution on the change of imaginary parts of parameters and boundary conditions in the neighborhood of zero. In other words, it was revealed that a small imaginary part may drive the real solution. It may be concluded that a small imaginary part, even if unobservable, is still an inherent characteristic of a physical quantity, being yet something like a hidden parameter, and manifests itself only indirectly forcing the system to move in this or that direction, which may be taken as a basis for experimental testing of the put forward hypothesis about a complex-valued nature of physical quantities.

# Appendix

## Table 1

```
                    RESULTS OF PROGRAM CPLSMN
ETA  = .1000000E+00
MU   = .5500000E-03
ALFA = .1000000E+03
REF0 = .1000000E-02
JMF0 = .1000000E-02
REDF0= .0000000E+00
JMDF0= .0000000E+00
REA0 = .1000000E-02
JMA0 = .0000000E+00
REDA0= .0000000E+00
JMDA0= .0000000E+00
NSTEP=  20
STEP = .1000000E+00
JTOL = 5
INCR = 0
X    =
CALCULATION WITH ACCURACY TOL= 1.0E-05
```

| # | X | ReF | JmF | ReA | JmA | dReF/dx | dReA/dx | ReNe | ReNi |
|---|---|---|---|---|---|---|---|---|---|
| 1. | 0.0000E+00 | 1.0000E-03 | 1.0000E-03 | 1.0000E-03 | 0.0000E+00 | 0.0000E+00 | 0.0000E+00 | 1.0001E+00 | 9.9900E-01 |
| 2. | 1.0000E-01 | -2.0486E-04 | 1.1003E-03 | 3.1204E-02 | 6.4996E-05 | -9.1797E-02 | 1.0804E+00 | 4.1619E-01 | 9.9972E-01 |
| 3. | 2.0000E-01 | -5.5666E-02 | 1.3607E-03 | 1.6956E-01 | 4.4236E-04 | -1.0614E+00 | 1.4235E+00 | 5.4582E-12 | 1.0423E+00 |
| 4. | 3.0000E-01 | -2.1578E-01 | 1.7663E-03 | 3.1307E-01 | 8.3340E-04 | -2.1639E+00 | 1.4499E+00 | 3.9106E-39 | 1.1821E+00 |
| 5. | 4.0000E-01 | -4.9536E-01 | 2.3868E-03 | 4.6034E-01 | 1.2275E-03 | -3.4765E+00 | 1.5005E+00 | 8.8542E-84 | 1.4776E+00 |
| 6. | 5.0000E-01 | -9.2525E-01 | 3.3763E-03 | 6.1458E-01 | 1.6228E-03 | -5.2225E+00 | 1.5941E+00 | 8.9690-149 | 2.0923E+00 |
| 7. | 6.0000E-01 | -1.5709E+00 | 5.1289E-03 | 7.8222E-01 | 2.0051E-03 | -7.9298E+00 | 1.7828E+00 | 1.3756-240 | 3.5538E+00 |
| 8. | 7.0000E-01 | -2.5959E+00 | 9.0065E-03 | 9.8011E-01 | 2.2913E-03 | -1.3333E+01 | 2.2597E+00 | 0.0000E+00 | 8.3336E+00 |
| 9. | 8.0000E-01 | -4.6296E+00 | 2.4540E-02 | 1.2818E+00 | 1.5602E-03 | -3.2695E+01 | 4.4761E+00 | 0.0000E+00 | 4.5421E+01 |
| 10. | 9.0000E-01 | -7.0966E+01 | 1.1411E+01 | 2.0536E+01 | -3.6290E+00 | -2.3907E+03 | 7.5581E+02 | 0.0000E+00 | 8.5097E-58 |
| 11. | 1.0000E+00 | -3.1004E+02 | 4.0077E+01 | 9.6117E+01 | -1.2771E+01 | -2.3907E+03 | 7.5581E+02 | 0.0000E+00 | 0.0000E+00 |
| 12. | 1.1000E+00 | -5.4911E+02 | 6.8743E+01 | 1.7170E+02 | -2.1914E+01 | -2.3907E+03 | 7.5581E+02 | 0.0000E+00 | 0.0000E+00 |
| 13. | 1.2000E+00 | -7.8819E+02 | 9.7410E+01 | 2.4728E+02 | -3.1056E+01 | -2.3907E+03 | 7.5581E+02 | 0.0000E+00 | 0.0000E+00 |
| 14. | 1.3000E+00 | -1.0273E+03 | 1.2608E+02 | 3.2286E+02 | -4.0199E+01 | -2.3907E+03 | 7.5581E+02 | 0.0000E+00 | 0.0000E+00 |
| 15. | 1.4000E+00 | -1.2663E+03 | 1.5474E+02 | 3.9844E+02 | -4.9341E+01 | -2.3907E+03 | 7.5581E+02 | 0.0000E+00 | 0.0000E+00 |
| 16. | 1.5000E+00 | -1.5054E+03 | 1.8341E+02 | 4.7402E+02 | -5.8484E+01 | -2.3907E+03 | 7.5581E+02 | 0.0000E+00 | 0.0000E+00 |
| 17. | 1.6000E+00 | -1.7445E+03 | 2.1207E+02 | 5.4960E+02 | -6.7626E+01 | -2.3907E+03 | 7.5581E+02 | 0.0000E+00 | 0.0000E+00 |
| 18. | 1.7000E+00 | -1.9836E+03 | 2.4074E+02 | 6.2518E+02 | -7.6769E+01 | -2.3907E+03 | 7.5581E+02 | 0.0000E+00 | 0.0000E+00 |
| 19. | 1.8000E+00 | -2.2226E+03 | 2.6941E+02 | 7.0076E+02 | -8.5911E+01 | -2.3907E+03 | 7.5581E+02 | 0.0000E+00 | 0.0000E+00 |
| 20. | 1.9000E+00 | -2.4617E+03 | 2.9807E+02 | 7.7634E+02 | -9.5054E+01 | -2.3907E+03 | 7.5581E+02 | 0.0000E+00 | 0.0000E+00 |
| 21. | 2.0000E+00 | -2.7008E+03 | 3.2674E+02 | 8.5192E+02 | -1.0420E+02 | -2.3907E+03 | 7.5581E+02 | 0.0000E+00 | 0.0000E+00 |

(Program CPLSMN)

**Table 2**

```
                    RESULTS OF PROGRAM CPLSMN
ETA  = .1000000E+00
MU   = .5500000E-03
ALFA = .1000000E+03
REF0 = .1000000E-02
JMF0 = .1000000E-01
REDF0= .0000000E+00
JMDF0= .0000000E+00
REA0 = .1000000E-02
JMA0 = .0000000E+00
REDA0= .0000000E+00
JMDA0= .0000000E+00
NSTEP=  20
STEP = .1000000E+00
JTOL =  5
INCR =  0
X    =
CALCULATION WITH ACCURACY TOL= 1.0E-05
```

| # | X | ReF | JmF | ReA | JmA | dReF/dx | dReA/dx | ReNe | ReNi |
|---|---|---|---|---|---|---|---|---|---|
| 1. | 0.0000E+00 | 1.0000E-3 | 1.0000E-02 | 1.0000E-03 | 0.0000E+00 | 0.0000E+0 | 0.0000E+0 | 1.0000E+00 | 9.9895E-01 |
| 2. | 1.0000E-01 | -2.0416E-4 | 1.1003E-02 | 3.1198E-02 | 6.4995E-04 | -9.1741E-2 | 1.0802E+0 | 4.1647E-01 | 9.9966E-01 |
| 3. | 2.0000E-01 | -5.5651E-2 | 1.3608E-02 | 1.6956E-01 | 4.4251E-03 | -1.0612E+0 | 1.4236E+0 | 5.5523E-12 | 1.0422E+00 |
| 4. | 3.0000E-01 | -2.1574E-1 | 1.7663E-02 | 3.1309E-01 | 8.3372E-03 | -2.1636E+0 | 1.4501E+0 | 4.1309E-39 | 1.1819E+00 |
| 5. | 4.0000E-01 | -4.9527E-1 | 2.3867E-02 | 4.6037E-01 | 1.2280E-02 | -3.4759E+0 | 1.5006E+0 | 9.9213E-84 | 1.4772E+00 |
| 6. | 5.0000E-01 | -9.2506E-1 | 3.3761E-02 | 6.1462E-01 | 1.6235E-02 | -5.2210E+0 | 1.5942E+0 | 1.0856-148 | 2.0909E+00 |
| 7. | 6.0000E-01 | -1.5705E+0 | 5.1284E-02 | 7.8228E-01 | 2.0060E-02 | -7.9254E+0 | 1.7830E+0 | 1.8146-240 | 3.5481E+00 |
| 8. | 7.0000E-01 | -2.5944E+0 | 9.0030E-02 | 9.8017E-01 | 2.2924E-02 | -1.3311E+1 | 2.2591E+0 | 0.0000E+00 | 8.2896E+00 |
| 9. | 8.0000E-01 | -4.6171E+0 | 2.4449E-01 | 1.2810E+00 | 1.5693E-02 | -3.2288E+1 | 4.4356E+0 | 0.0000E+00 | 4.3585E+01 |
| 10. | 9.0000E-01 | -6.5430E+0 | 6.5722E+00 | 1.3652E+00 | -7.8301E-01 | 9.0910E+1 | -1.1537E+1 | 0.0000E+00 | 2.8566E+02 |
| 11. | 1.0000E+00 | -2.3843E+0 | 6.9383E+00 | 9.1691E-01 | -6.5634E-01 | 2.6178E+1 | -2.7704E+0 | -3.2955-162 | 8.7453E+00 |
| 12. | 1.1000E+00 | 1.1030E-2 | 7.3371E+00 | 6.5469E-01 | -6.5040E-01 | 2.3046E+1 | -2.5860E+0 | 3.2486E-03 | 4.8736E-01 |
| 13. | 1.2000E+00 | 2.3067E+0 | 7.8223E+00 | 7.9836E-01 | -5.4137E-01 | 2.2927E+1 | 1.5267E+0 | 8.2399-136 | 2.6598E-03 |
| 14. | 1.3000E+00 | 4.5996E+0 | 8.3206E+00 | 9.5091E-01 | -4.3078E-01 | 2.2930E+1 | 1.5248E+0 | -5.3194-280 | -3.1698E-03 |
| 15. | 1.4000E+00 | 6.8927E+0 | 8.8201E+00 | 1.1034E+00 | -3.2028E-01 | 2.2932E+1 | 1.5246E+0 | 0.0000E+00 | -4.8090E-04 |
| 16. | 1.5000E+00 | 9.1859E+0 | 9.3196E+00 | 1.2558E+00 | -2.0977E-01 | 2.2932E+1 | 1.5246E+0 | 0.0000E+00 | -4.7706E-05 |
| 17. | 1.6000E+00 | 1.1479E+1 | 9.8191E+00 | 1.4083E+00 | -9.9274E-02 | 2.2932E+1 | 1.5246E+0 | 0.0000E+00 | -3.5950E-06 |
| 18. | 1.7000E+00 | 1.3772E+1 | 1.0319E+01 | 1.5607E+00 | 1.1227E-02 | 2.2932E+1 | 1.5246E+0 | 0.0000E+00 | -1.9583E-07 |
| 19. | 1.8000E+00 | 1.6066E+1 | 1.0818E+01 | 1.7132E+00 | 1.2173E-01 | 2.2932E+1 | 1.5246E+0 | 0.0000E+00 | -4.3806E-09 |
| 20. | 1.9000E+00 | 1.8359E+1 | 1.1318E+01 | 1.8657E+00 | 2.3223E-01 | 2.2932E+1 | 1.5246E+0 | 0.0000E+00 | 6.1755E-10 |
| 21. | 2.0000E+00 | 2.0652E+1 | 1.1817E+01 | 2.0181E+00 | 3.4273E-01 | 2.2932E+1 | 1.5246E+0 | 0.0000E+00 | 1.4668E-10 |

(Program CPLSMN)

## Table 3

RESULTS OF PROGRAM COLD

ETA = .5000000E+00
GAMMA = .2500000E+00
KSI0 =-.2000000E+02
NSTEP = 350
STEP = .1000000E+00
JTOL = 5
INCR =
CULCULATION WITH ACCURACY TOL= 1.0E-05

| # | X | F | A | dF/dx | dA/dx | Ne | Ni | Te | Ti |
|---|---|---|---|---|---|---|---|---|---|
| 1 | -2.0000E+01 | 2.5764E-10 | 4.5400E-05 | 2.5764E-10 | 0.0000E+00 | 1.0000E+00 | 1.0000E+00 | 3.4050E-05 | 1.1350E-05 |
| 50 | -1.5100E+01 | 8.6338E-09 | 2.6503E-04 | 8.5719E-09 | 1.3056E-04 | 1.0000E+00 | 1.0000E+00 | 1.9878E-04 | 6.6259E-05 |
| 100 | -1.0100E+01 | 1.2847E-06 | 3.2052E-03 | 1.2837E-06 | 1.6024E-03 | 1.0000E+00 | 1.0000E+00 | 2.4039E-03 | 8.0130E-04 |
| 150 | -5.1000E+00 | 1.9067E-04 | 3.9048E-02 | 1.9063E-04 | 1.9525E-02 | 1.0003E+00 | 1.0001E+00 | 2.9294E-02 | 9.7629E-03 |
| 200 | -1.0000E-01 | 2.8898E-02 | 4.7786E-01 | 2.9513E-02 | 2.4114E-01 | 1.0455E+00 | 1.0147E+00 | 3.7469E-01 | 1.2122E-01 |
| 210 | 9.0000E-01 | 8.1667E-02 | 7.9439E-01 | 8.6938E-02 | 4.0799E-01 | 1.1414E+00 | 1.0427E+00 | 6.8006E-01 | 2.0708E-01 |
| 220 | 1.9000E+00 | 2.5344E-01 | 1.3454E+00 | 3.1812E-01 | 7.3971E-01 | 1.6621E+00 | 1.1470E+00 | 1.6771E+00 | 3.8578E-01 |
| 221 | 2.0000E+00 | 2.8806E-01 | 1.4221E+00 | 3.7692E-01 | 7.9504E-01 | 1.8425E+00 | 1.1708E+00 | 1.9651E+00 | 4.1622E-01 |
| 222 | 2.1000E+00 | 3.2949E-01 | 1.5047E+00 | 4.5577E-01 | 8.6005E-01 | 2.1290E+00 | 1.2006E+00 | 2.4027E+00 | 4.5165E-01 |
| 223 | 2.2000E+00 | 3.8039E-01 | 1.5946E+00 | 5.7046E-01 | 9.4065E-01 | 2.6697E+00 | 1.2395E+00 | 3.1928E+00 | 4.9413E-01 |
| 224 | 2.3000E+00 | 4.4630E-01 | 1.6939E+00 | 7.7159E-01 | 1.0548E+00 | 4.2474E+00 | 1.2937E+00 | 5.3961E+00 | 5.4784E-01 |
| 225 | 2.4000E+00 | 5.5010E-01 | 1.8109E+00 | 1.2653E+00 | 1.2748E+00 | 0.0000E+00 | 1.3868E+00 | 0.0000E+00 | 6.2784E-01 |
| 226 | 2.5000E+00 | 6.6948E-01 | 1.9392E+00 | 1.1200E+00 | 1.2918E+00 | 0.0000E+00 | 1.5246E+00 | 0.0000E+00 | 7.3914E-01 |
| 227 | 2.6000E+00 | 7.7358E-01 | 2.0694E+00 | 9.5911E-01 | 1.3120E+00 | 0.0000E+00 | 1.7011E+00 | 0.0000E+00 | 8.8008E-01 |
| 228 | 2.7000E+00 | 8.6062E-01 | 2.2018E+00 | 7.7783E-01 | 1.3362E+00 | 0.0000E+00 | 1.9364E+00 | 0.0000E+00 | 1.0659E+00 |
| 229 | 2.8000E+00 | 9.2822E-01 | 2.3368E+00 | 5.6867E-01 | 1.3659E+00 | 0.0000E+00 | 2.2669E+00 | 0.0000E+00 | 1.3243E+00 |
| 230 | 2.9000E+00 | 9.7300E-01 | 2.4752E+00 | 3.1880E-01 | 1.4035E+00 | 0.0000E+00 | 2.7673E+00 | 0.0000E+00 | 1.7124E+00 |
| 231 | 3.0000E+00 | 9.8983E-01 | 2.6179E+00 | 3.9356E-03 | 1.4536E+00 | 0.0000E+00 | 3.6099E+00 | 0.0000E+00 | 2.3626E+00 |
| 232 | 3.1000E+00 | 9.6992E-01 | 2.7667E+00 | -4.2856E-01 | 1.5265E+00 | 0.0000E+00 | 5.2248E+00 | 0.0000E+00 | 3.6138E+00 |
| 233 | 3.2000E+00 | 8.9675E-01 | 2.9247E+00 | -1.0773E+00 | 1.6421E+00 | 0.0000E+00 | 7.6668E+00 | 0.0000E+00 | 5.6057E+00 |
| 234 | 3.3000E+00 | 7.5107E-01 | 3.0960E+00 | -1.8104E+00 | 1.7798E+00 | 0.0000E+00 | 6.2739E+00 | 0.0000E+00 | 4.8560E+00 |
| 235 | 3.4000E+00 | 5.4266E-01 | 3.2793E+00 | -2.3235E+00 | 1.8818E+00 | 0.0000E+00 | 4.2054E+00 | 0.0000E+00 | 3.4477E+00 |
| 236 | 3.5000E+00 | 2.9134E-01 | 3.4715E+00 | -2.6854E+00 | 1.9580E+00 | 0.0000E+00 | 3.1445E+00 | 0.0000E+00 | 2.7290E+00 |
| 237 | 3.6000E+00 | 8.1981E-03 | 3.6705E+00 | -2.9675E+00 | 2.0209E+00 | 0.0000E+00 | 2.5494E+00 | 0.0000E+00 | 2.3394E+00 |
| 238 | 3.7000E+00 | -3.0061E-01 | 3.8754E+00 | -3.2023E+00 | 2.0763E+00 | 0.0000E+00 | 2.1738E+00 | 0.0000E+00 | 2.1061E+00 |
| 239 | 3.8000E+00 | -6.3124E-01 | 4.0856E+00 | -3.4061E+00 | 2.1269E+00 | 0.0000E+00 | 1.9161E+00 | 0.0000E+00 | 1.9572E+00 |
| 240 | 3.9000E+00 | -9.8110E-01 | 4.3007E+00 | -3.5878E+00 | 2.1745E+00 | 0.0000E+00 | 1.7289E+00 | 0.0000E+00 | 1.8589E+00 |
| 241 | 4.0000E+00 | -1.3483E+00 | 4.5204E+00 | -3.7532E+00 | 2.2201E+00 | 0.0000E+00 | 1.5871E+00 | 0.0000E+00 | 1.7936E+00 |
| 242 | 4.1000E+00 | -1.7313E+00 | 4.7447E+00 | -3.9062E+00 | 2.2644E+00 | 0.0000E+00 | 1.4766E+00 | 0.0000E+00 | 1.7515E+00 |
| 243 | 4.2000E+00 | -2.1292E+00 | 4.9733E+00 | -4.0493E+00 | 2.3078E+00 | 0.0000E+00 | 1.3884E+00 | 0.0000E+00 | 1.7263E+00 |
| 244 | 4.3000E+00 | -2.5409E+00 | 5.2062E+00 | -4.1844E+00 | 2.3508E+00 | 0.0000E+00 | 1.3171E+00 | 0.0000E+00 | 1.7143E+00 |
| 250 | 4.9000E+00 | -5.2711E+00 | 6.6946E+00 | -4.8942E+00 | 2.6132E+00 | 0.0000E+00 | 1.0947E+00 | 0.0000E+00 | 1.8321E+00 |
| 260 | 5.9000E+00 | -1.0707E+01 | 9.5657E+00 | -6.0005E+00 | 3.1743E+00 | 0.0000E+00 | 1.2552E+00 | 0.0000E+00 | 3.0017E+00 |
| 261 | 6.0000E+00 | -1.1314E+01 | 9.8870E+00 | -6.1306E+00 | 3.2534E+00 | 0.0000E+00 | 1.3517E+00 | 0.0000E+00 | 3.3411E+00 |
| 262 | 6.1000E+00 | -1.1934E+01 | 1.0217E+01 | -6.2728E+00 | 3.3427E+00 | 0.0000E+00 | 1.5023E+00 | 0.0000E+00 | 3.8373E+00 |
| 263 | 6.2000E+00 | -1.2569E+01 | 1.0556E+01 | -6.4350E+00 | 3.4481E+00 | 0.0000E+00 | 1.7677E+00 | 0.0000E+00 | 4.6649E+00 |
| 264 | 6.3000E+00 | -1.3222E+01 | 1.0907E+01 | -6.6376E+00 | 3.5841E+00 | 0.0000E+00 | 2.3876E+00 | 0.0000E+00 | 6.5105E+00 |
| 265 | 6.4000E+00 | -1.3901E+01 | 1.1276E+01 | -7.0484E+00 | 3.8700E+00 | 0.0000E+00 | 1.7404E+01 | 0.0000E+00 | 4.9064E+01 |
| 266 | 6.5000E+00 | -1.4612E+01 | 1.1668E+01 | -7.1089E+00 | 3.9126E+00 | 0.0000E+00 | 0.0000E+00 | 0.0000E+00 | 0.0000E+00 |
| 267 | 6.6000E+00 | -1.5323E+01 | 1.2059E+01 | -7.1089E+00 | 3.9126E+00 | 0.0000E+00 | 0.0000E+00 | 0.0000E+00 | 0.0000E+00 |
| 268 | 6.7000E+00 | -1.6034E+01 | 1.2450E+01 | -7.1089E+00 | 3.9126E+00 | 0.0000E+00 | 0.0000E+00 | 0.0000E+00 | 0.0000E+00 |
| 269 | 6.8000E+00 | -1.6745E+01 | 1.2841E+01 | -7.1089E+00 | 3.9126E+00 | 0.0000E+00 | 0.0000E+00 | 0.0000E+00 | 0.0000E+00 |
| 270 | 6.9000E+00 | -1.7456E+01 | 1.3233E+01 | -7.1089E+00 | 3.9126E+00 | 0.0000E+00 | 0.0000E+00 | 0.0000E+00 | 0.0000E+00 |
| 351 | 1.5000E+01 | -7.5038E+01 | 4.4925E+01 | -7.1089E+00 | 3.9126E+00 | 0.0000E+00 | 0.0000E+00 | 0.0000E+00 | 0.0000E+00 |

(Program COLD)

**Table 4**

RESULTS OF PROGRAM CCOLD

ETA = .5000000E+00
GAMMA = .2500000E+00
KSI0 =-.2000000E+02
REF0 = .2576400E-09
JMF0 = .1000000E-14
REA0 = .4540000E-04
JMA0 = .0000000E+00
REDF0 = .2576400E-09
JMDF0 = .0000000E+00
REDA0 = .0000000E+00
JMDA0 = .0000000E+00
NSTEP = 999
STEP = .1000000E+00
JTOL = 5
INCR = 0
CULCULATION WITH ACCURACY TOL= 1.0E-05

| # | X | ReF | JmF | ReA | JmA | dReF/dx | dReA/dx | ReNe | ReNi |
|---|---|---|---|---|---|---|---|---|---|
| 1. | -2.0000E+01 | 2.5764E-10 | 1.0000E-15 | 4.5400E-05 | 0.0000E+00 | 2.5764E-10 | 0.0000E+00 | 1.0000E+00 | 1.0000E+00 |
| 50. | -1.5100E+01 | 8.6339E-09 | 1.0000E-15 | 2.6503E-04 | 0.0000E+00 | 8.5719E-09 | 1.3056E-04 | 1.0000E+00 | 1.0000E+00 |
| 100. | -1.0100E+01 | 1.2847E-06 | 1.0000E-15 | 3.2052E-03 | 0.0000E+00 | 1.2837E-06 | 1.6024E-03 | 1.0000E+00 | 1.0000E+00 |
| 150. | -5.1000E+00 | 1.9067E-04 | 1.0000E-15 | 3.9048E-02 | 0.0000E+00 | 1.9064E-04 | 1.9525E-02 | 1.0003E+00 | 1.0001E+00 |
| 200. | -1.0000E-01 | 2.8898E-02 | 1.0000E-15 | 4.7786E-01 | 0.0000E+00 | 2.9513E-02 | 2.4114E-01 | 1.0455E+00 | 1.0147E+00 |
| 210. | 9.0000E-01 | 8.1668E-02 | 1.0000E-15 | 7.9439E-01 | 0.0000E+00 | 8.6939E-02 | 4.0799E-01 | 1.1414E+00 | 1.0427E+00 |
| 220. | 1.9000E+00 | 2.5344E-01 | 1.0000E-15 | 1.3454E+00 | 0.0000E+00 | 3.1812E-01 | 7.3971E-01 | 1.6621E+00 | 1.1470E+00 |
| 221. | 2.0000E+00 | 2.8807E-01 | 1.0000E-15 | 1.4221E+00 | 0.0000E+00 | 3.7692E-01 | 7.9504E-01 | 1.8425E+00 | 1.1708E+00 |
| 222. | 2.1000E+00 | 3.2949E-01 | 1.0000E-15 | 1.5047E+00 | 0.0000E+00 | 4.5577E-01 | 8.6005E-01 | 2.1290E+00 | 1.2006E+00 |
| 223. | 2.2000E+00 | 3.8039E-01 | 1.0000E-15 | 1.5946E+00 | 0.0000E+00 | 5.7047E-01 | 9.4065E-01 | 2.6697E+00 | 1.2395E+00 |
| 224. | 2.3000E+00 | 4.4630E-01 | 1.0000E-15 | 1.6939E+00 | 0.0000E+00 | 7.7160E-01 | 1.0548E+00 | 4.2475E+00 | 1.2937E+00 |
| 225. | 2.4000E+00 | 5.6057E-01 | -3.6604E-11 | 1.8144E+00 | -1.2182E-11 | 1.7896E+00 | 1.4504E+00 | 9.5333E+00 | 1.3949E+00 |
| 226. | 2.5000E+00 | 7.6949E-01 | -2.4362E-10 | 1.9733E+00 | -8.2569E-11 | 2.3212E+00 | 1.7091E+00 | 5.2586E+00 | 1.6398E+00 |
| 227. | 2.6000E+00 | 1.0155E+00 | -5.7352E-10 | 2.1541E+00 | -2.0062E-10 | 2.5593E+00 | 1.8977E+00 | 3.3959E+00 | 2.2237E+00 |
| 228. | 2.7000E+00 | 1.2722E+00 | -9.4272E-10 | 2.3519E+00 | -3.5508E-10 | 2.4938E+00 | 2.0623E+00 | 2.2185E+00 | 7.4203E+00 |
| 229. | 2.8000E+00 | 1.4903E+00 | 1.6366E-06 | 2.5686E+00 | -2.5060E-07 | 2.0727E+00 | 2.2353E+00 | 1.4495E+00 | 2.5198E+00 |
| 230. | 2.9000E+00 | 1.6932E+00 | 3.4711E-06 | 2.7972E+00 | -5.5973E-07 | 1.9921E+00 | 2.3320E+00 | 1.0774E+00 | 1.7262E+00 |
| 231. | 3.0000E+00 | 1.8894E+00 | 5.3025E-06 | 3.0345E+00 | -9.0034E-07 | 1.9344E+00 | 2.4124E+00 | 8.6208E-01 | 1.3865E+00 |
| 232. | 3.1000E+00 | 2.0803E+00 | 7.1295E-06 | 3.2794E+00 | -1.2692E-06 | 1.8852E+00 | 2.4841E+00 | 7.2002E-01 | 1.1849E+00 |
| 233. | 3.2000E+00 | 2.2666E+00 | 8.9531E-06 | 3.5312E+00 | -1.6636E-06 | 1.8407E+00 | 2.5503E+00 | 6.1851E-01 | 1.0468E+00 |
| 234. | 3.3000E+00 | 2.4486E+00 | 1.0774E-05 | 3.7893E+00 | -2.0814E-06 | 1.7993E+00 | 2.6126E+00 | 5.4200E-01 | 9.4419E-01 |
| 235. | 3.4000E+00 | 2.6265E+00 | 1.2594E-05 | 4.0536E+00 | -2.5212E-06 | 1.7601E+00 | 2.6723E+00 | 4.8207E-01 | 8.6379E-01 |
| 236. | 3.5000E+00 | 2.8006E+00 | 1.4413E-05 | 4.3237E+00 | -2.9817E-06 | 1.7228E+00 | 2.7298E+00 | 4.3377E-01 | 7.9841E-01 |
| 237. | 3.6000E+00 | 2.9711E+00 | 1.6232E-05 | 4.5995E+00 | -3.4619E-06 | 1.6871E+00 | 2.7859E+00 | 3.9395E-01 | 7.4375E-01 |
| 238. | 3.7000E+00 | 3.1381E+00 | 1.8050E-05 | 4.8809E+00 | -3.9611E-06 | 1.6528E+00 | 2.8406E+00 | 3.6053E-01 | 6.9707E-01 |
| 239. | 3.8000E+00 | 3.3017E+00 | 1.9868E-05 | 5.1676E+00 | -4.4785E-06 | 1.6198E+00 | 2.8944E+00 | 3.3207E-01 | 6.5655E-01 |
| 240. | 3.9000E+00 | 3.4621E+00 | 2.1686E-05 | 5.4597E+00 | -5.0137E-06 | 1.5879E+00 | 2.9474E+00 | 3.0752E-01 | 6.2088E-01 |
| 241. | 4.0000E+00 | 3.6193E+00 | 2.3503E-05 | 5.7571E+00 | -5.5662E-06 | 1.5571E+00 | 2.9998E+00 | 2.8613E-01 | 5.8913E-01 |
| 242. | 4.1000E+00 | 3.7735E+00 | 2.5320E-05 | 6.0597E+00 | -6.1356E-06 | 1.5273E+00 | 3.0516E+00 | 2.6731E-01 | 5.6062E-01 |
| 243. | 4.2000E+00 | 3.9248E+00 | 2.7136E-05 | 6.3674E+00 | -6.7215E-06 | 1.4984E+00 | 3.1030E+00 | 2.5064E-01 | 5.3480E-01 |
| 244. | 4.3000E+00 | 4.0733E+00 | 2.8952E-05 | 6.6803E+00 | -7.3237E-06 | 1.4705E+00 | 3.1540E+00 | 2.3575E-01 | 5.1126E-01 |
| 245. | 4.4000E+00 | 4.2189E+00 | 3.0767E-05 | 6.9982E+00 | -7.9419E-06 | 1.4433E+00 | 3.2048E+00 | 2.2239E-01 | 4.8968E-01 |
| 246. | 4.5000E+00 | 4.3619E+00 | 3.2580E-05 | 7.3212E+00 | -8.5759E-06 | 1.4170E+00 | 3.2552E+00 | 2.1032E-01 | 4.6979E-01 |
| 247. | 4.6000E+00 | 4.5024E+00 | 3.4393E-05 | 7.6493E+00 | -9.2253E-06 | 1.3914E+00 | 3.3055E+00 | 1.9937E-01 | 4.5138E-01 |
| 248. | 4.7000E+00 | 4.6402E+00 | 3.6204E-05 | 7.9823E+00 | -9.8901E-06 | 1.3666E+00 | 3.3556E+00 | 1.8939E-01 | 4.3427E-01 |
| 249. | 4.8000E+00 | 4.7757E+00 | 3.8014E-05 | 8.3204E+00 | -1.0570E-05 | 1.3424E+00 | 3.4055E+00 | 1.8025E-01 | 4.1832E-01 |
| 250. | 4.9000E+00 | 4.9087E+00 | 3.9822E-05 | 8.6634E+00 | -1.1265E-05 | 1.3190E+00 | 3.4553E+00 | 1.7186E-01 | 4.0339E-01 |
| 260. | 5.9000E+00 | 6.1216E+00 | 5.7788E-05 | 1.2366E+01 | -1.9009E-05 | 1.1156E+00 | 3.9498E+00 | 1.1483E-01 | 2.9337E-01 |
| 270. | 6.9000E+00 | 7.1547E+00 | 7.5538E-05 | 1.6563E+01 | -2.8171E-05 | 9.5666E-01 | 4.4428E+00 | 8.3749E-02 | 2.2517E-01 |
| 300. | 9.9000E+00 | 9.5024E+00 | 1.2864E-04 | 3.2115E+01 | -6.5477E-05 | 6.3900E-01 | 5.9273E+00 | 4.2074E-02 | 1.2108E-01 |
| 400. | 1.9900E+01 | 1.3332E+01 | 3.1055E-04 | 1.1629E+02 | -3.3032E-04 | 2.1376E-01 | 1.0913E+01 | 1.1481E-02 | 3.4282E-02 |
| 500. | 2.9900E+01 | 1.4592E+01 | 4.9866E-04 | 2.5040E+02 | -8.7560E-04 | 5.7883E-02 | 1.5910E+01 | 5.3265E-03 | 1.5961E-02 |
| 600. | 3.9900E+01 | 1.4731E+01 | 6.9051E-04 | 4.3450E+02 | -1.7691E-03 | -2.3066E-02 | 2.0909E+01 | 3.0690E-03 | 9.2035E-03 |
| 700. | 4.9900E+01 | 1.4235E+01 | 8.8489E-04 | 6.6859E+02 | -3.0690E-03 | -7.2610E-02 | 2.5909E+01 | 1.9943E-03 | 5.9821E-03 |
| 800. | 5.9900E+01 | 1.3333E+01 | 1.0811E-03 | 9.5268E+02 | -4.8274E-03 | -1.0621E-01 | 3.0909E+01 | 1.3996E-03 | 4.1985E-03 |
| 900. | 6.9900E+01 | 1.2145E+01 | 1.2788E-03 | 1.2868E+03 | -7.0922E-03 | -1.3037E-01 | 3.5909E+01 | 1.0362E-03 | 3.1085E-03 |
| 1000. | 7.9900E+01 | 1.0747E+01 | 1.4776E-03 | 1.6709E+03 | -9.9078E-03 | -1.4857E-01 | 4.0909E+01 | 7.9800E-04 | 2.3939E-03 |

(Program CCOLD)

**Table 5**

RESULTS OF PROGRAM CCOLD

ETA = .5000000E+00
GAMMA = .2500000E+00
KSI0 =-.2000000E+02
REF0 = .2576400E-09
JMF0 = .1000000E-12
REA0 = .4540000E-04
JMA0 = .0000000E+00
REDF0 = .2576400E-09
JMDF0 = .0000000E+00
REDA0 = .0000000E+00
JMDA0 = .0000000E+00
NSTEP = 999
STEP = .1000000E+00
JTOL = 5
INCR = 0
CULCULATION WITH ACCURACY TOL= 1.0E-05

| | X | dReF/dx | dJmF/dx | dReA/dx | dJmA/dx | ReNe | JmNe | ReNi | JmNi |
|---|---|---|---|---|---|---|---|---|---|
| 1. | -2.0000E+01 | 2.5764E-10 | 0.0000E+00 | 0.0000E+00 | 0.0000E+00 | 1.0000E+00 | 0.0000E+00 | 1.0000E+00 | 0.0000E+00 |
| 50. | -1.5100E+01 | 8.5719E-09 | 0.0000E+00 | 1.3056E-04 | 0.0000E+00 | 1.0000E+00 | 0.0000E+00 | 1.0000E+00 | 0.0000E+00 |
| 100. | -1.0100E+01 | 1.2837E-06 | 0.0000E+00 | 1.6024E-03 | 0.0000E+00 | 1.0000E+00 | 0.0000E+00 | 1.0000E+00 | 0.0000E+00 |
| 150. | -5.1000E+00 | 1.9064E-04 | 0.0000E+00 | 1.9525E-02 | 0.0000E+00 | 1.0003E+00 | 0.0000E+00 | 1.0001E+00 | 0.0000E+00 |
| 200. | -1.0000E-01 | 2.9513E-02 | 0.0000E+00 | 2.4114E-01 | 0.0000E+00 | 1.0455E+00 | 0.0000E+00 | 1.0147E+00 | 0.0000E+00 |
| 210. | 9.0000E-01 | 8.6939E-02 | 0.0000E+00 | 4.0799E-01 | 0.0000E+00 | 1.1414E+00 | 0.0000E+00 | 1.0427E+00 | 0.0000E+00 |
| 220. | 1.9000E+00 | 3.1812E-01 | 0.0000E+00 | 7.3971E-01 | 0.0000E+00 | 1.6621E+00 | 0.0000E+00 | 1.1470E+00 | 0.0000E+00 |
| 221. | 2.0000E+00 | 3.7692E-01 | 0.0000E+00 | 7.9504E-01 | 0.0000E+00 | 1.8425E+00 | 0.0000E+00 | 1.1708E+00 | 0.0000E+00 |
| 222. | 2.1000E+00 | 4.5577E-01 | 0.0000E+00 | 8.6005E-01 | 0.0000E+00 | 2.1290E+00 | 0.0000E+00 | 1.2006E+00 | 0.0000E+00 |
| 223. | 2.2000E+00 | 5.7047E-01 | 0.0000E+00 | 9.4065E-01 | 0.0000E+00 | 2.6697E+00 | 0.0000E+00 | 1.2395E+00 | 0.0000E+00 |
| 224. | 2.3000E+00 | 7.7160E-01 | 0.0000E+00 | 1.0548E+00 | 0.0000E+00 | 4.2475E+00 | 0.0000E+00 | 1.2937E+00 | 0.0000E+00 |
| 225. | 2.4000E+00 | 1.7896E+00 | -1.3496E-07 | 1.4504E+00 | -4.5050E-08 | 9.5333E+00 | 1.2559E-06 | 1.3949E+00 | 2.7558E-09 |
| 226. | 2.5000E+00 | 2.3212E+00 | -2.6466E-07 | 1.7091E+00 | -9.2225E-08 | 5.2586E+00 | 1.2733E-06 | 1.6398E+00 | 3.0237E-08 |
| 227. | 2.6000E+00 | 2.5593E+00 | -3.5674E-07 | 1.8977E+00 | -1.3314E-07 | 3.3959E+00 | 7.0026E-07 | 2.2237E+00 | 1.8038E-07 |
| 228. | 2.7000E+00 | 2.4938E+00 | -2.5987E-07 | 2.0623E+00 | -1.7463E-07 | 2.2185E+00 | 2.4635E-07 | 7.4203E+00 | 1.1313E-05 |
| 229. | 2.8000E+00 | 2.0730E+00 | 3.9438E-04 | 2.2352E+00 | -6.3084E-05 | 1.4496E+00 | 1.0422E-04 | 2.5196E+00 | 1.2659E-04 |
| 230. | 2.9000E+00 | 1.9924E+00 | 3.9449E-04 | 2.3319E+00 | -7.0184E-05 | 1.0775E+00 | 9.3831E-05 | 1.7261E+00 | 8.5048E-05 |
| 231. | 3.0000E+00 | 1.9348E+00 | 3.9348E-04 | 2.4124E+00 | -7.6600E-05 | 8.6214E-01 | 7.6069E-05 | 1.3864E+00 | 6.6127E-05 |
| 232. | 3.1000E+00 | 1.8856E+00 | 3.9260E-04 | 2.4841E+00 | -8.2364E-05 | 7.2007E-01 | 6.1802E-05 | 1.1848E+00 | 5.4406E-05 |
| 233. | 3.2000E+00 | 1.8411E+00 | 3.9199E-04 | 2.5502E+00 | -8.7622E-05 | 6.1855E-01 | 5.1054E-05 | 1.0467E+00 | 4.6117E-05 |
| 234. | 3.3000E+00 | 1.7997E+00 | 3.9159E-04 | 2.6126E+00 | -9.2496E-05 | 5.4203E-01 | 4.2921E-05 | 9.4415E-01 | 3.9802E-05 |
| 235. | 3.4000E+00 | 1.7605E+00 | 3.9134E-04 | 2.6722E+00 | -9.7073E-05 | 4.8210E-01 | 3.6660E-05 | 8.6376E-01 | 3.4757E-05 |
| 236. | 3.5000E+00 | 1.7232E+00 | 3.9120E-04 | 2.7298E+00 | -1.0142E-04 | 4.3379E-01 | 3.1748E-05 | 7.9838E-01 | 3.0592E-05 |
| 237. | 3.6000E+00 | 1.6876E+00 | 3.9110E-04 | 2.7858E+00 | -1.0557E-04 | 3.9397E-01 | 2.7825E-05 | 7.4372E-01 | 2.7072E-05 |
| 238. | 3.7000E+00 | 1.6533E+00 | 3.9104E-04 | 2.8406E+00 | -1.0958E-04 | 3.6055E-01 | 2.4641E-05 | 6.9705E-01 | 2.4043E-05 |
| 239. | 3.8000E+00 | 1.6202E+00 | 3.9098E-04 | 2.8944E+00 | -1.1345E-04 | 3.3209E-01 | 2.2019E-05 | 6.5652E-01 | 2.1402E-05 |
| 240. | 3.9000E+00 | 1.5883E+00 | 3.9091E-04 | 2.9474E+00 | -1.1723E-04 | 3.0754E-01 | 1.9832E-05 | 6.2086E-01 | 1.9075E-05 |
| 241. | 4.0000E+00 | 1.5575E+00 | 3.9083E-04 | 2.9997E+00 | -1.2091E-04 | 2.8614E-01 | 1.7988E-05 | 5.8911E-01 | 1.7008E-05 |
| 242. | 4.1000E+00 | 1.5277E+00 | 3.9072E-04 | 3.0516E+00 | -1.2451E-04 | 2.6733E-01 | 1.6417E-05 | 5.6060E-01 | 1.5158E-05 |
| 243. | 4.2000E+00 | 1.4989E+00 | 3.9058E-04 | 3.1030E+00 | -1.2804E-04 | 2.5065E-01 | 1.5067E-05 | 5.3478E-01 | 1.3494E-05 |
| 244. | 4.3000E+00 | 1.4709E+00 | 3.9040E-04 | 3.1540E+00 | -1.3151E-04 | 2.3576E-01 | 1.3896E-05 | 5.1124E-01 | 1.1992E-05 |
| 245. | 4.4000E+00 | 1.4438E+00 | 3.9020E-04 | 3.2047E+00 | -1.3493E-04 | 2.2239E-01 | 1.2873E-05 | 4.8966E-01 | 1.0629E-05 |
| 250. | 4.9000E+00 | 1.3194E+00 | 3.8867E-04 | 3.4553E+00 | -1.5135E-04 | 1.7187E-01 | 9.2692E-06 | 4.0338E-01 | 5.4233E-06 |
| 260. | 5.9000E+00 | 1.1161E+00 | 3.8370E-04 | 3.9498E+00 | -1.8198E-04 | 1.1483E-01 | 5.7269E-06 | 2.9337E-01 | 1.5041E-07 |
| 270. | 6.9000E+00 | 9.5718E-01 | 3.8042E-04 | 4.4427E+00 | -2.1349E-04 | 8.3751E-02 | 4.0386E-06 | 2.2517E-01 | 2.5386E-06 |
| 300. | 9.9000E+00 | 6.3956E-01 | 3.8249E-04 | 5.9272E+00 | -3.2675E-04 | 4.2075E-02 | 2.0534E-06 | 1.2108E-01 | 3.8000E-06 |
| 400. | 1.9900E+01 | 2.1429E-01 | 3.9933E-04 | 1.0912E+01 | -8.4634E-04 | 1.1481E-02 | 7.1254E-07 | 3.4282E-02 | 2.0178E-06 |
| 500. | 2.9900E+01 | 5.8396E-02 | 4.0946E-04 | 1.5910E+01 | -1.5285E-03 | 5.3266E-03 | 4.0299E-07 | 1.5962E-02 | 1.1903E-06 |
| 600. | 3.9900E+01 | -2.2554E-02 | 4.1594E-04 | 2.0909E+01 | -2.3445E-03 | 3.0690E-03 | 2.6975E-07 | 9.2036E-03 | 8.0448E-07 |
| 700. | 4.9900E+01 | -7.2100E-02 | 4.2052E-04 | 2.5909E+01 | -3.2784E-03 | 1.9943E-03 | 1.9747E-07 | 5.9821E-03 | 5.9079E-07 |
| 800. | 5.9900E+01 | -1.0562E-01 | 4.2399E-04 | 3.0909E+01 | -4.3197E-03 | 1.3996E-03 | 1.5294E-07 | 4.1985E-03 | 4.5813E-07 |
| 900. | 6.9900E+01 | -1.2989E-01 | 4.2673E-04 | 3.5909E+01 | -5.4603E-03 | 1.0362E-03 | 1.2314E-07 | 3.1085E-03 | 3.6910E-07 |
| 1000. | 7.9900E+01 | -1.4814E-01 | 4.2897E-04 | 4.0909E+01 | -6.6938E-03 | 7.9800E-04 | 1.0202E-07 | 2.3940E-03 | 3.0590E-07 |

(Program CCOLD)

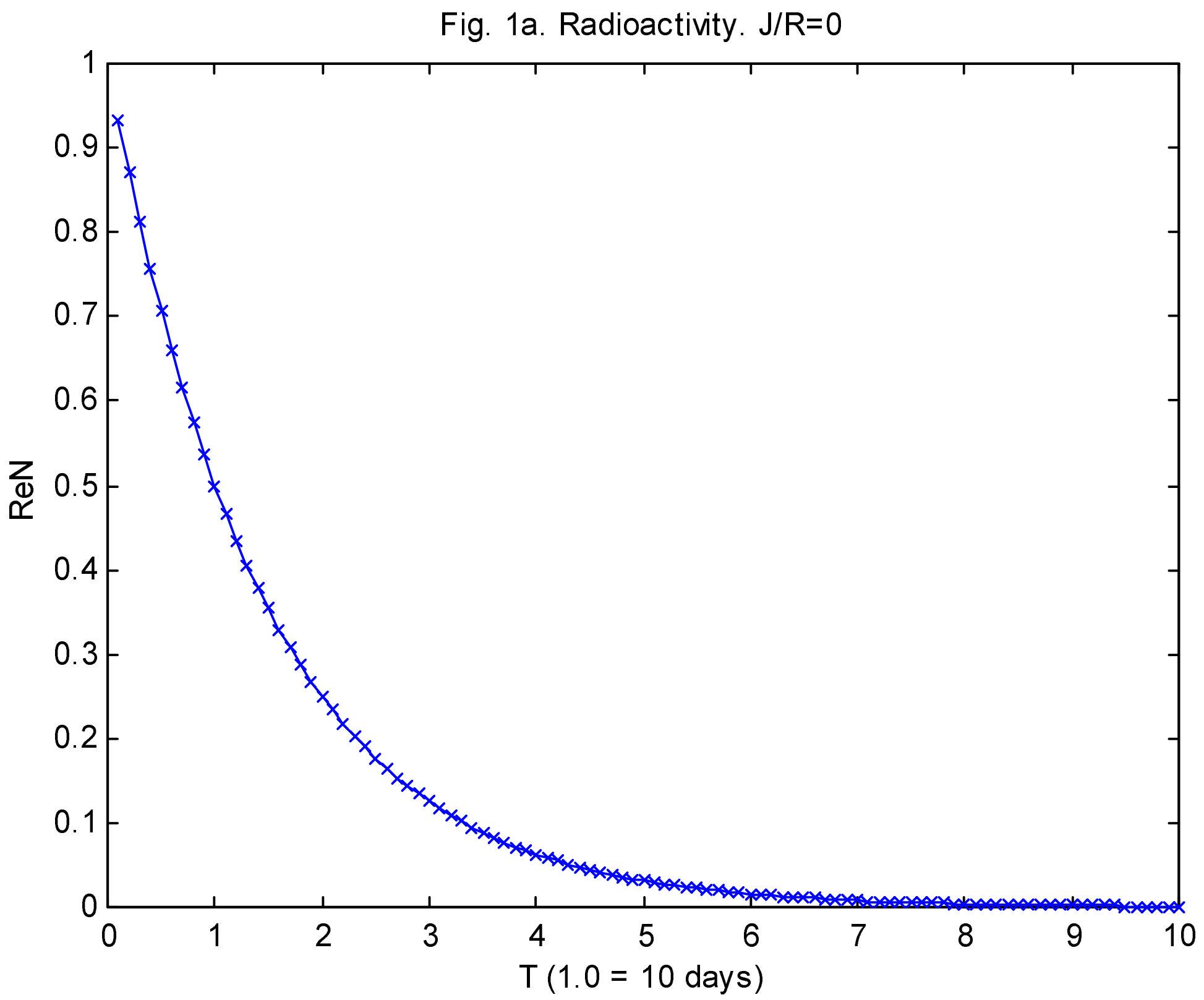
Fig. 1a. Radioactivity. J/R=0
ReN
T (1.0 = 10 days)
0
0.1
0.2
0.3
0.4
0.5
0.6
0.7
0.8
0.9
1
0 1 2 3 4 5 6 7 8 9 10

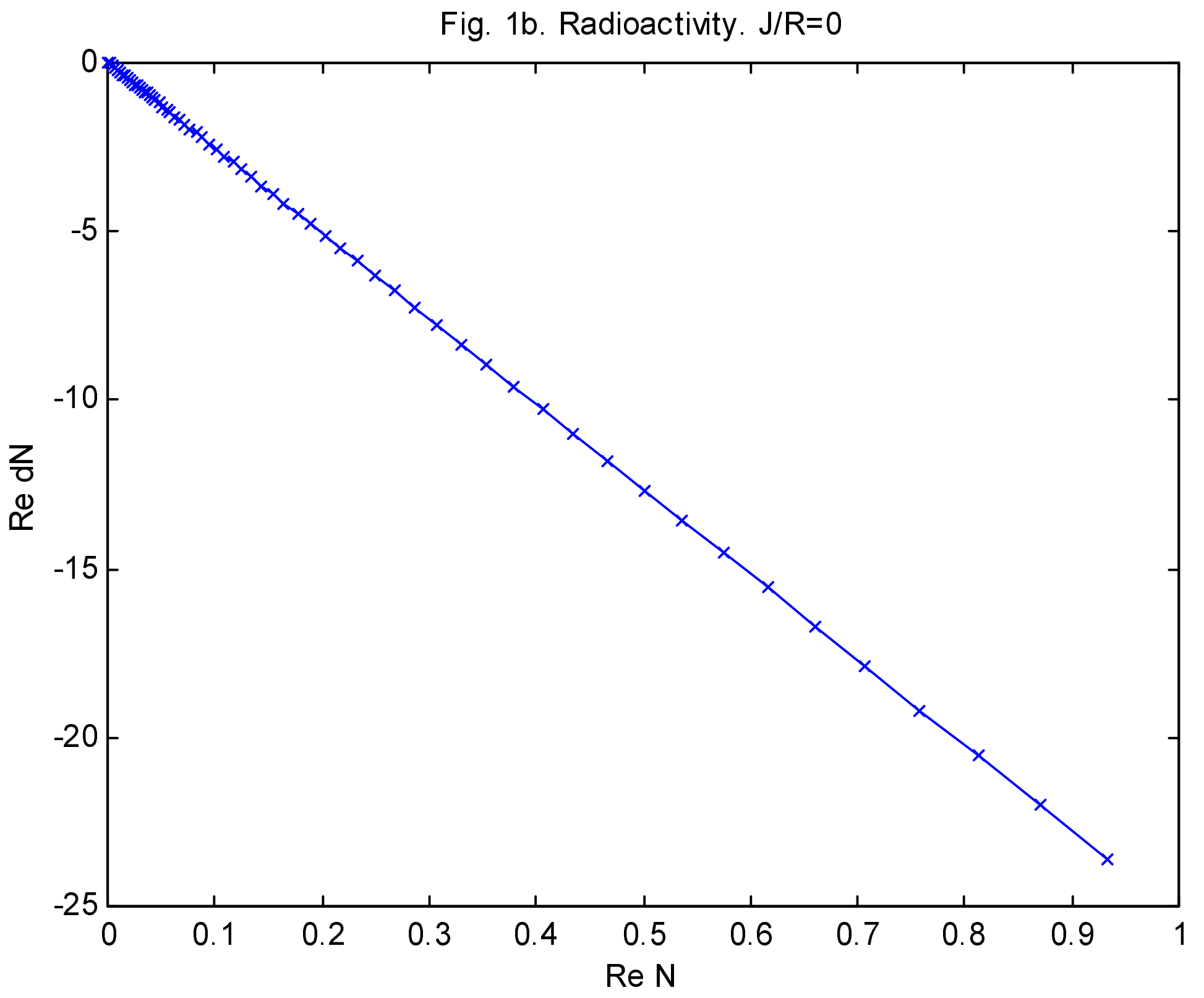
Fig. 1b. Radioactivity. J/R=0
Re dN
Re N
0
-5
-10
-15
-20
-25
0 0.1 0.2 0.3 0.4 0.5 0.6 0.7 0.8 0.9 1

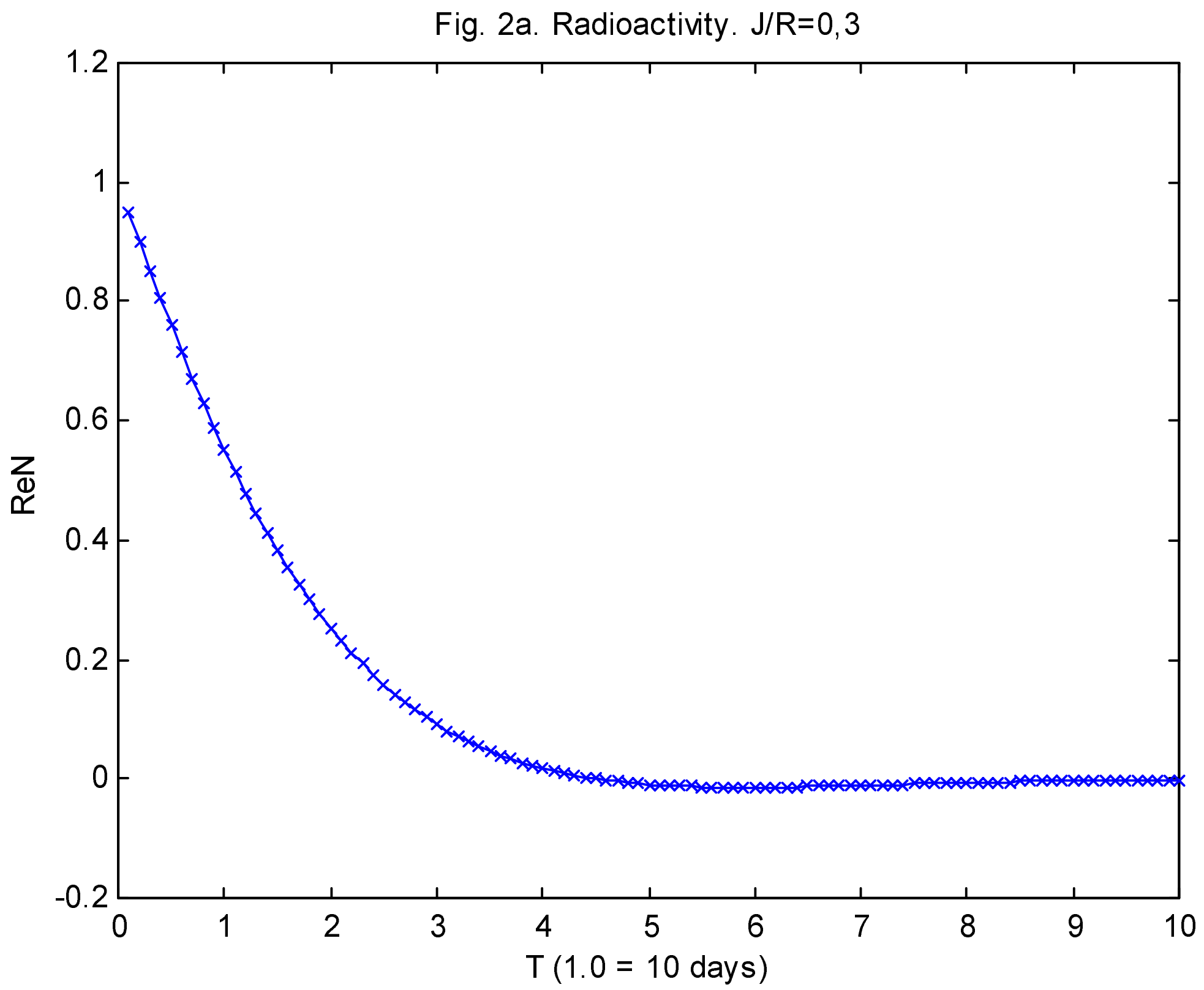

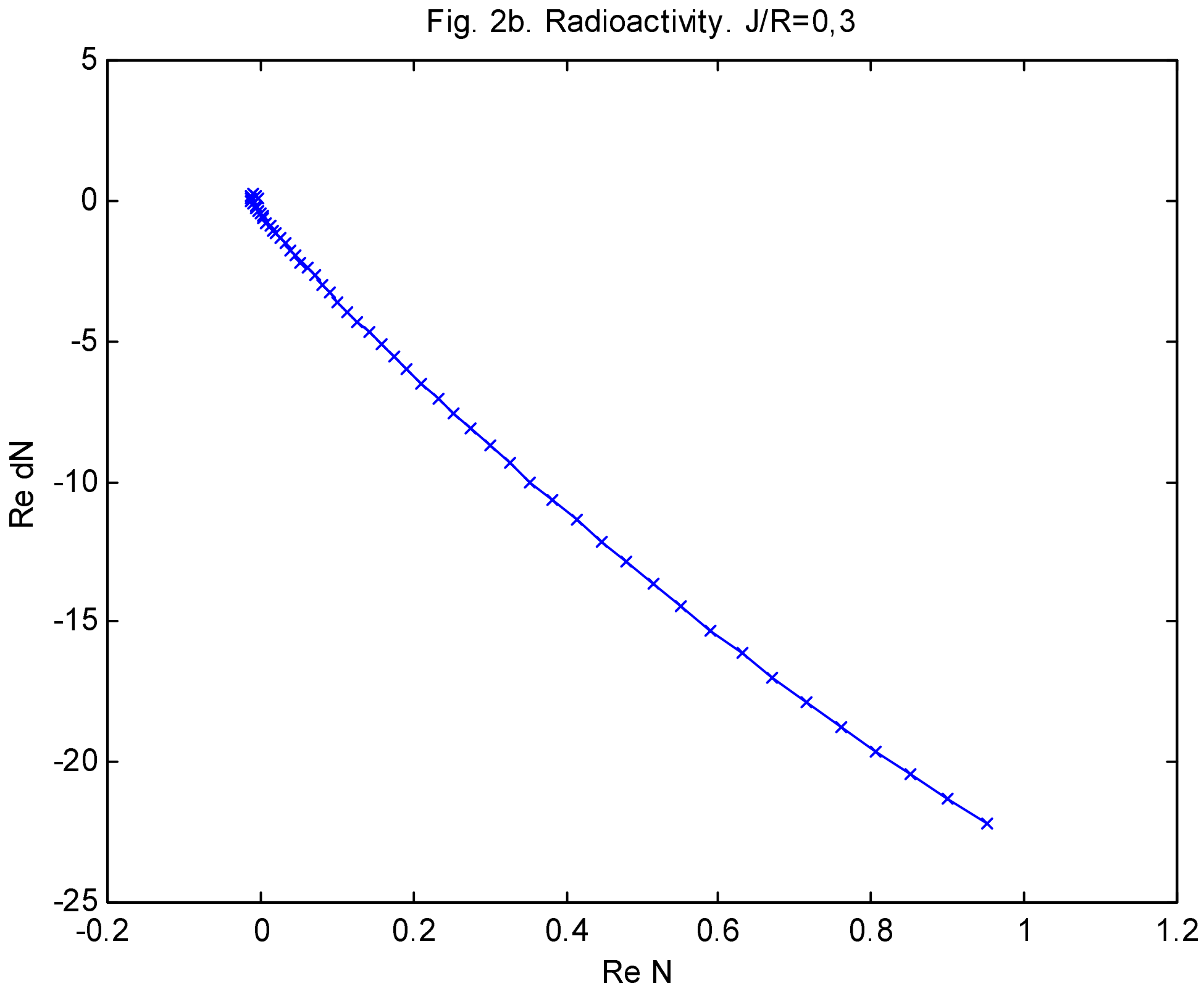

Fig. 3. Radioactivity. J/R=0,9

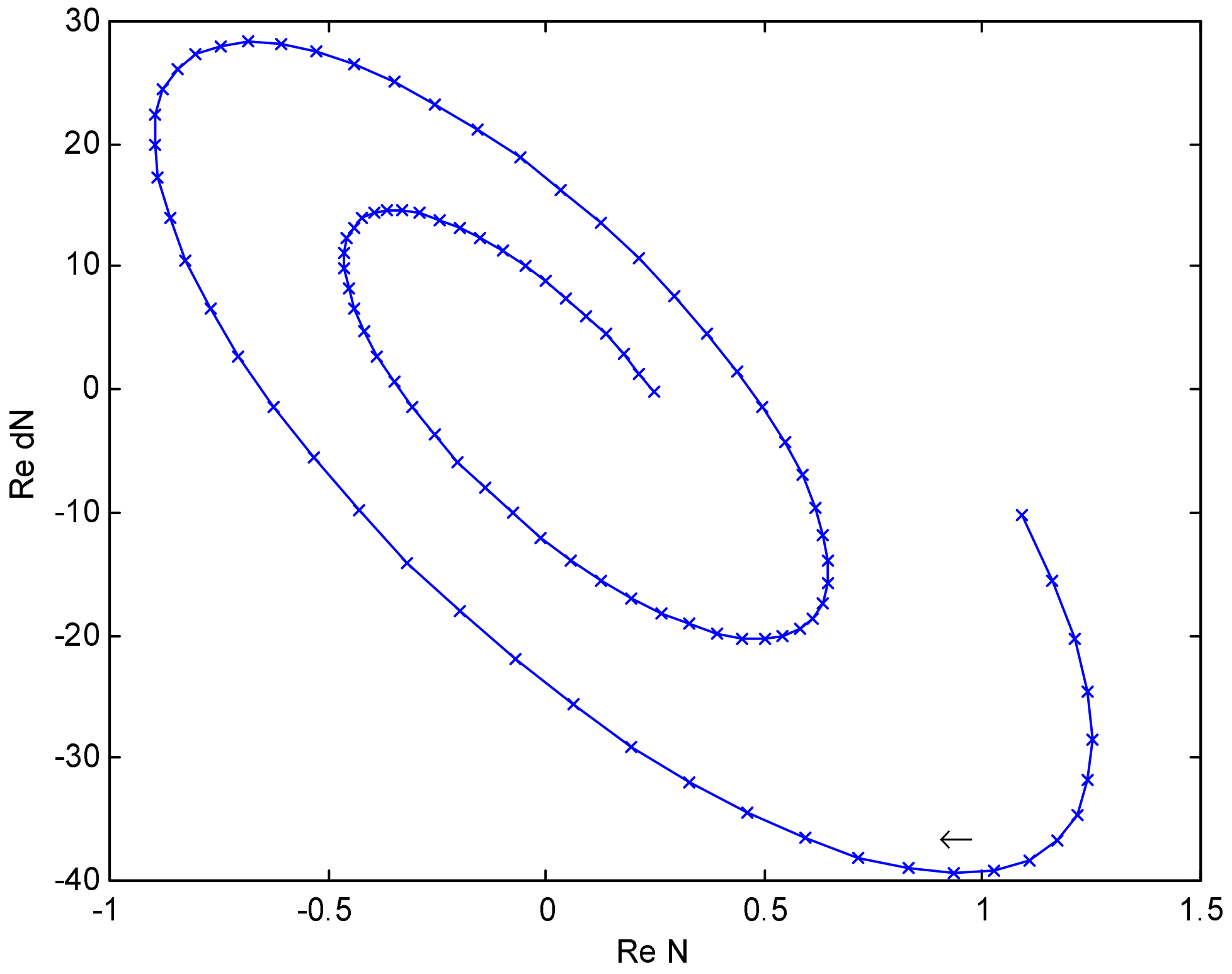

Fig. 4. Radioactivity. J/R=1,0

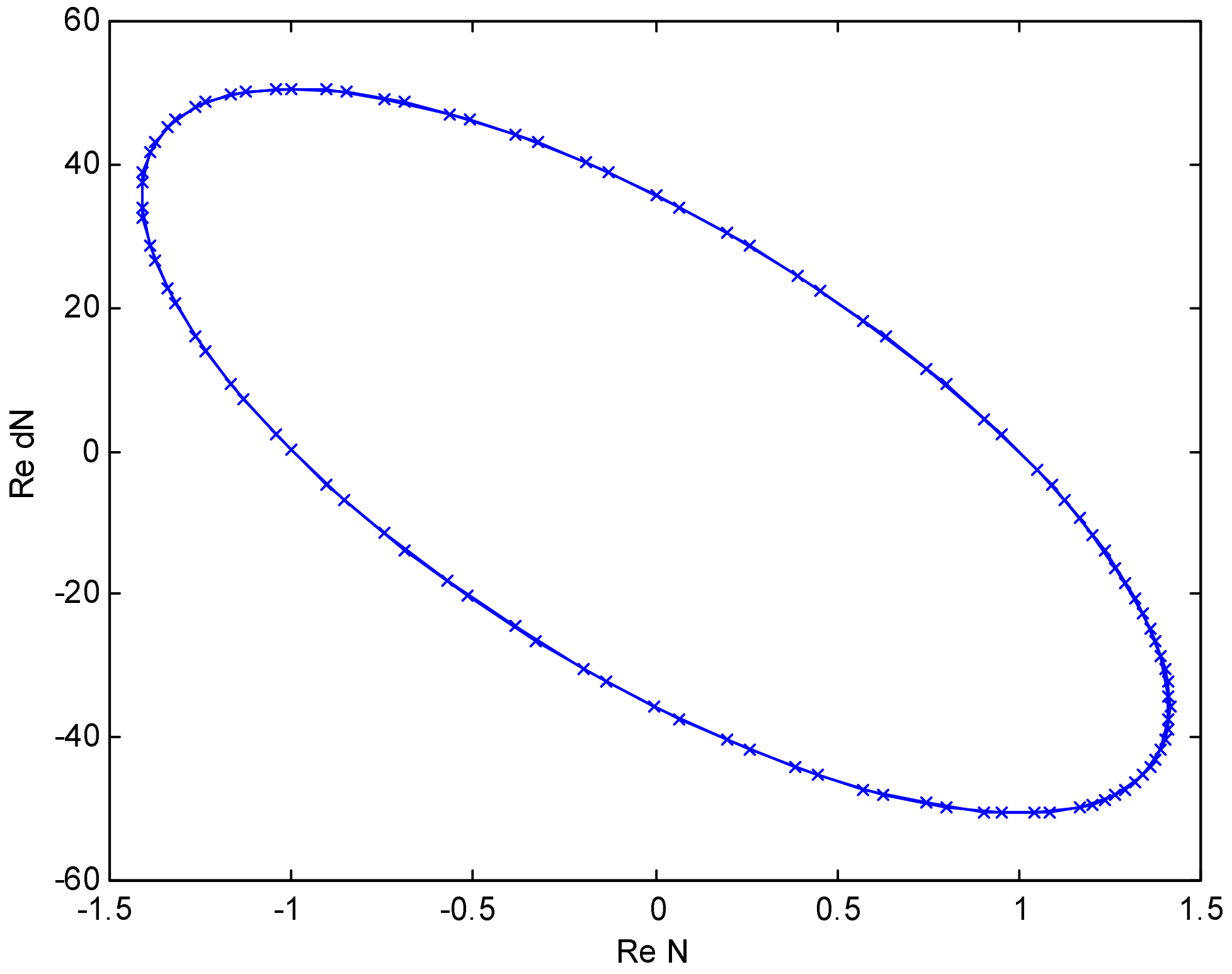

Fig. 5. Radioactivity. J/R=1,01

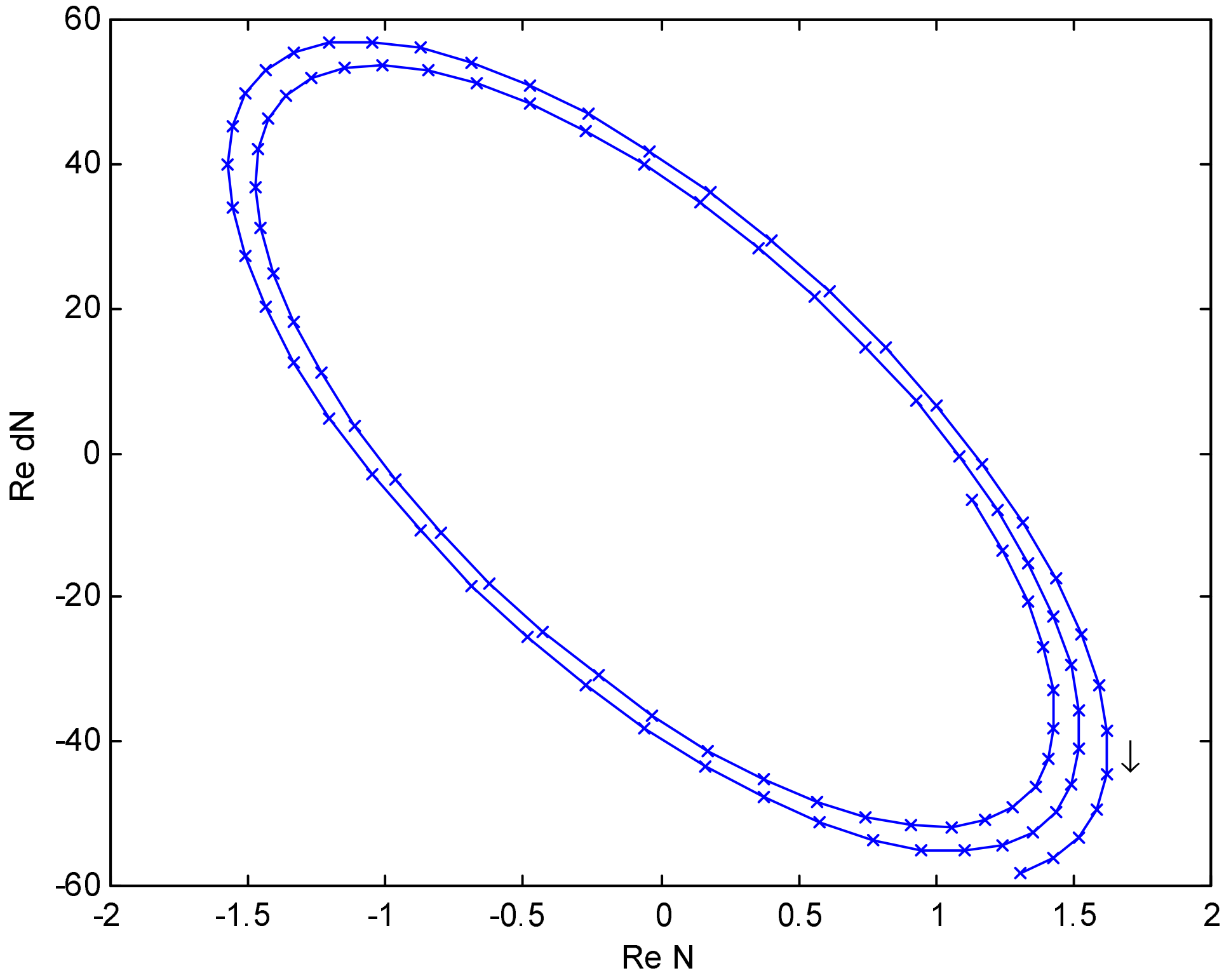

Fig. 6a . Oscillator. Phase-plane portrait

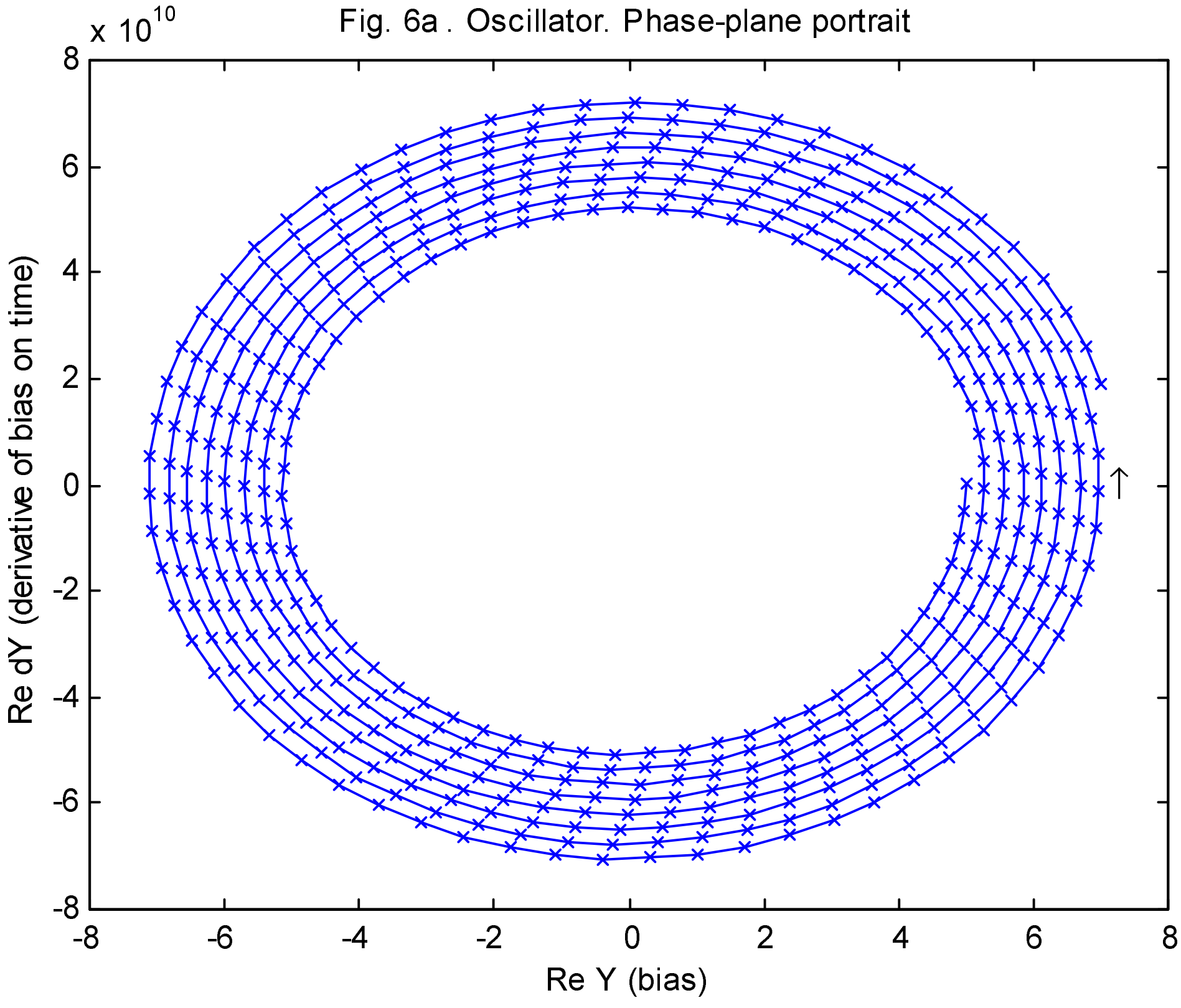

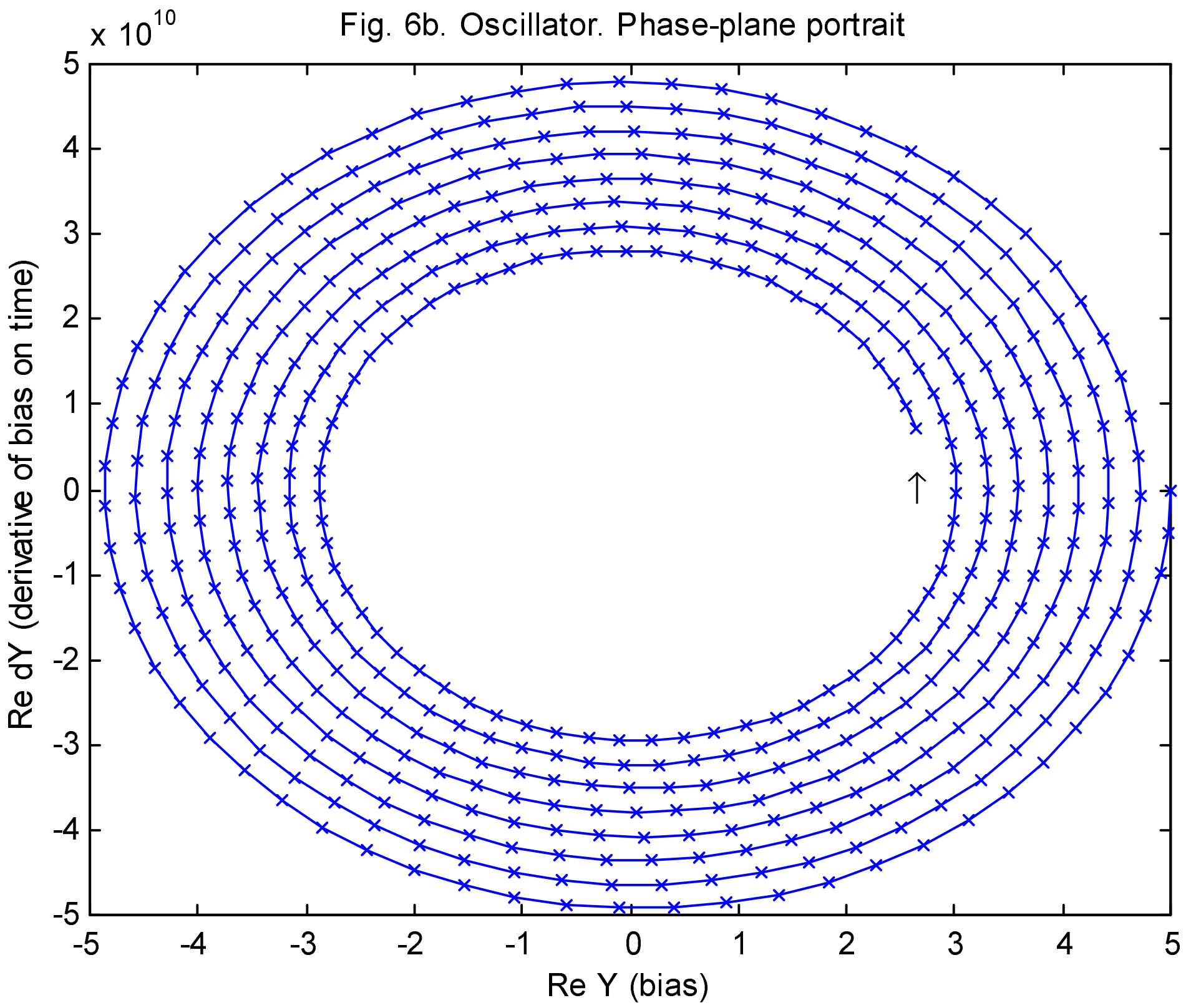
x 10^10
Fig. 6b. Oscillator. Phase-plane portrait
Re dY (derivative of bias on time)
Re Y (bias)
5
4
3
2
1
0
-1
-2
-3
-4
-5
-5 -4 -3 -2 -1 0 1 2 3 4 5

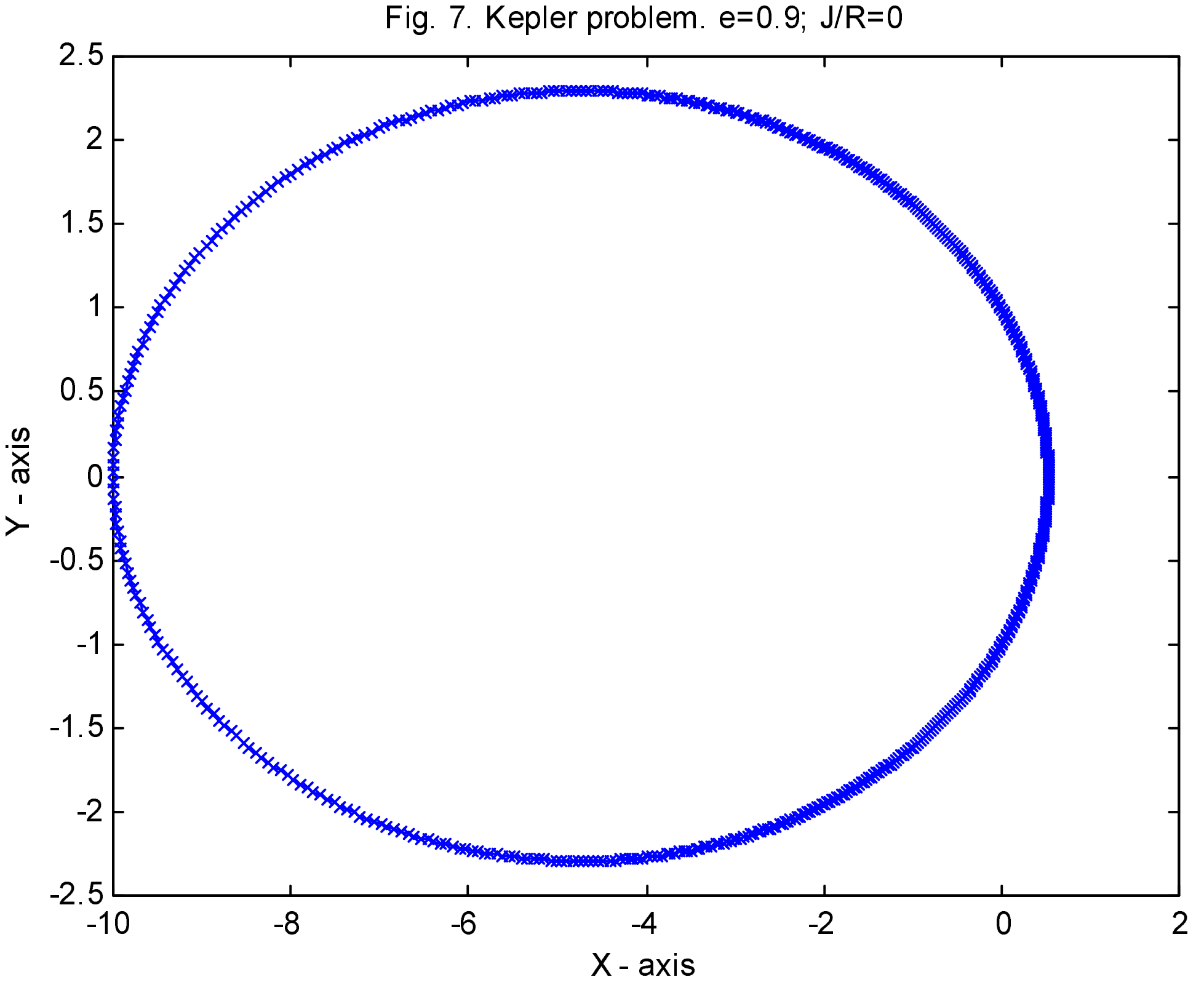

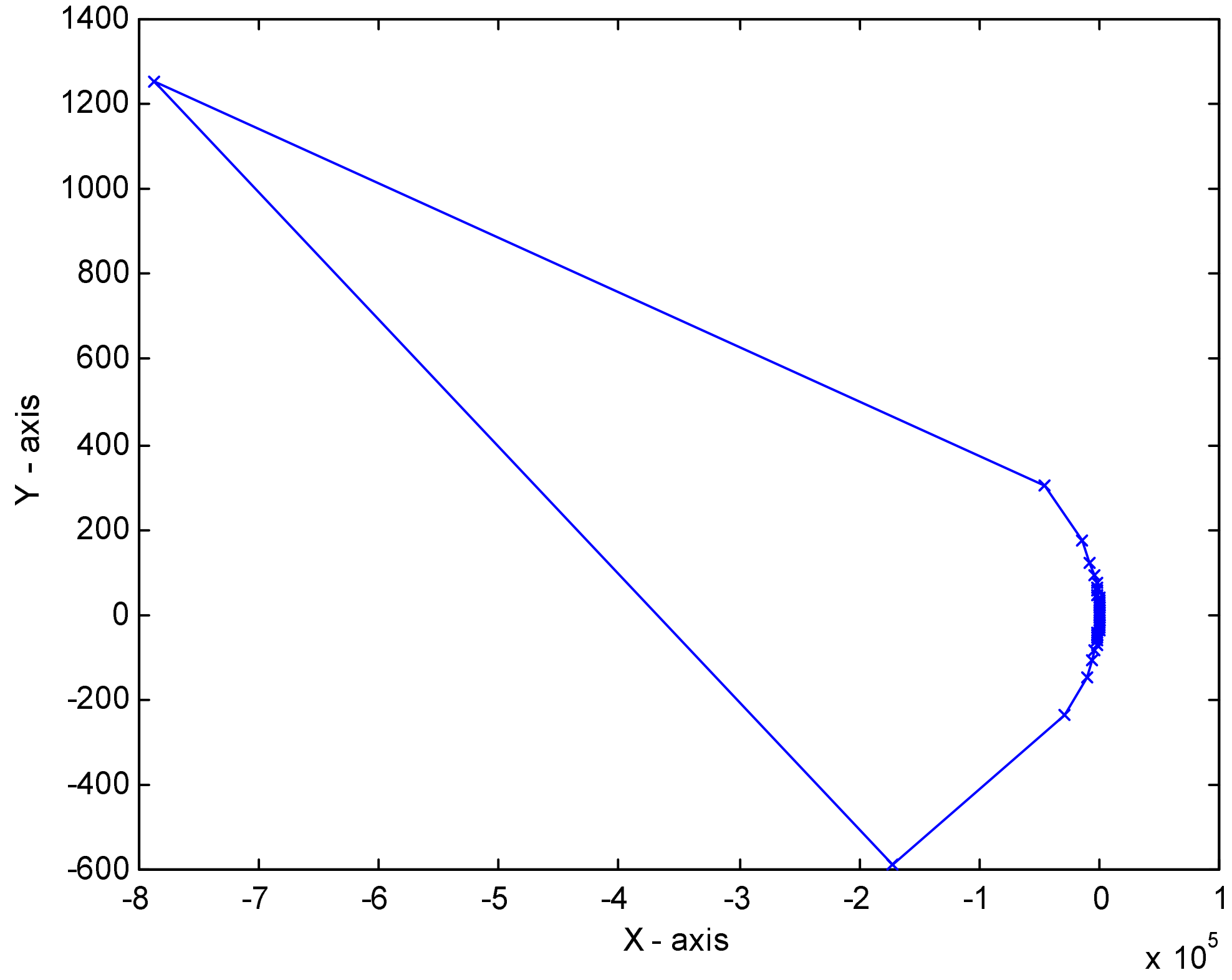

Fig. 8. Kepler problem. e=1.0; J/R=0

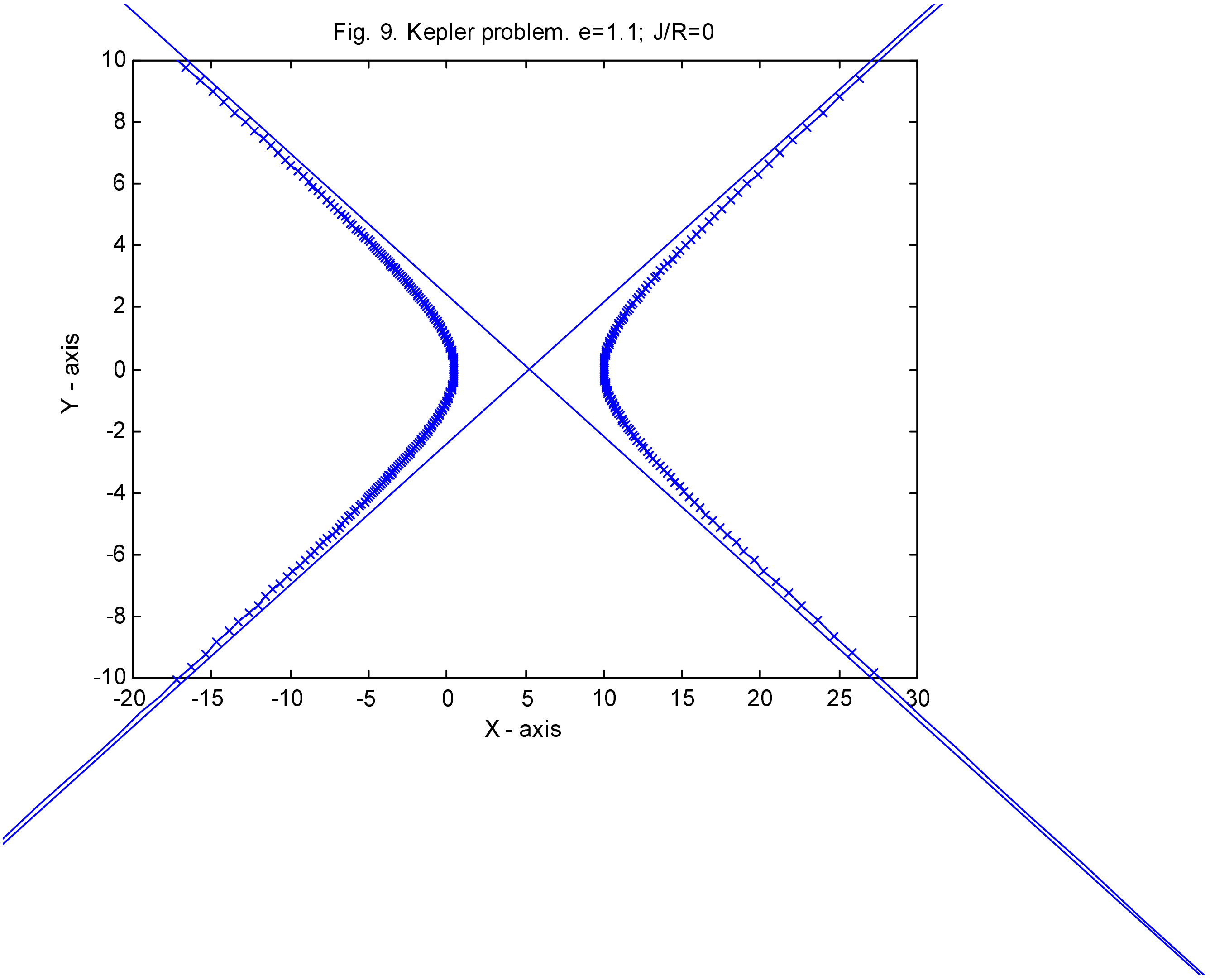

Fig. 9. Kepler problem. e=1.1; J/R=0

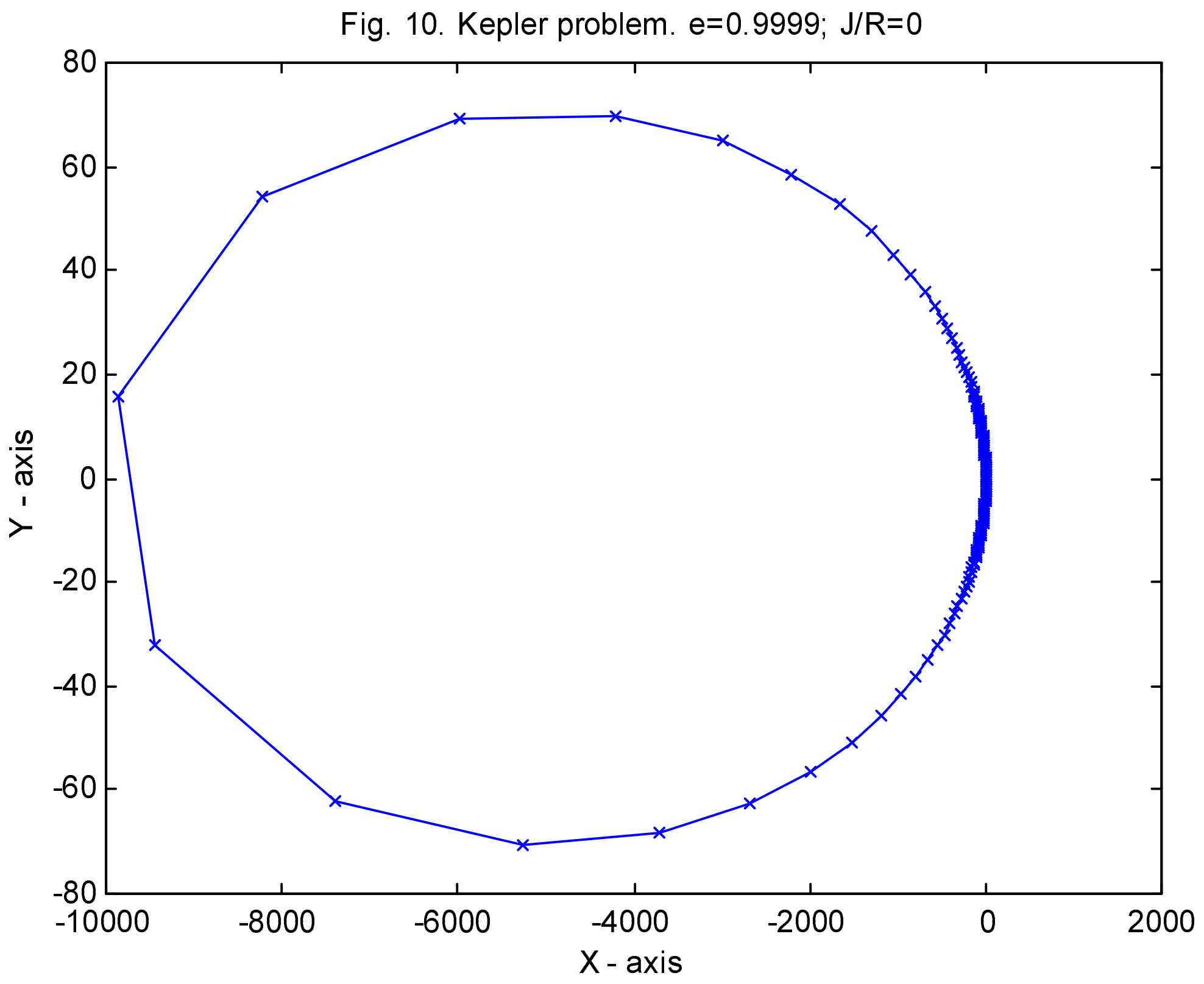

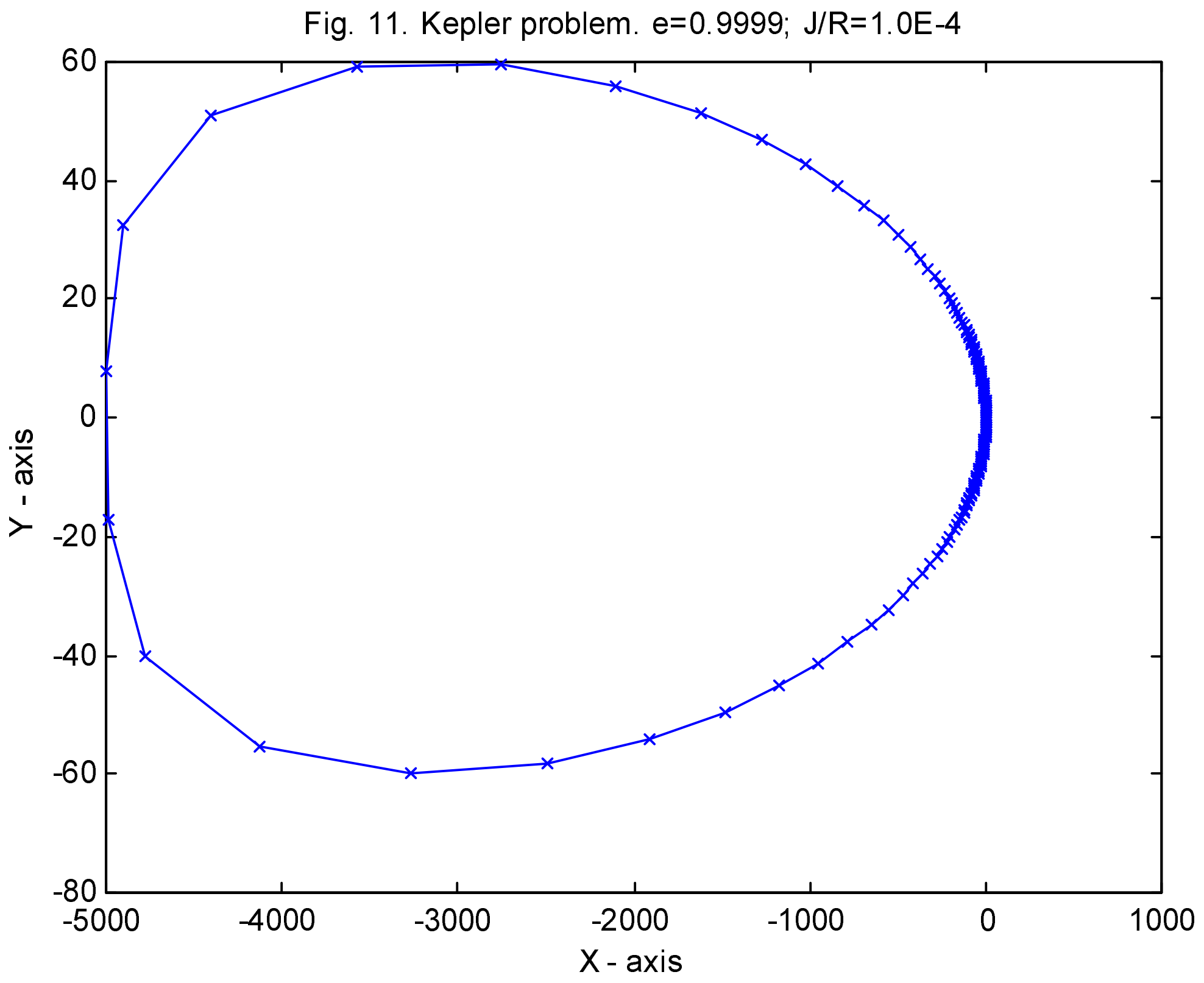

Fig. 12. Kepler problem. e=0.9999; J/R=1.0E-2

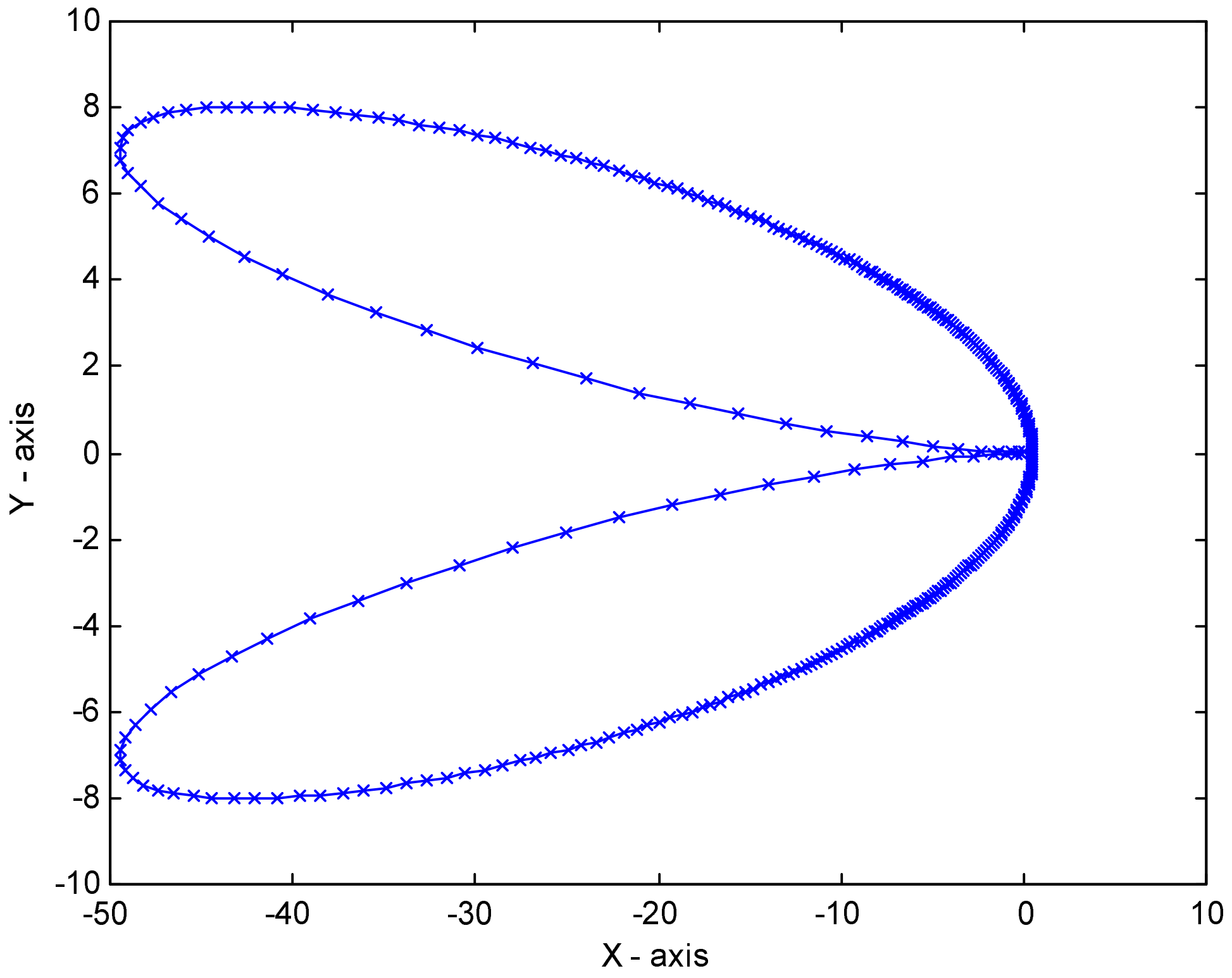

Fig. 13. Profiles of the real parts of electrical and magnetic fields. Im$\psi$(0)=1.0E-3

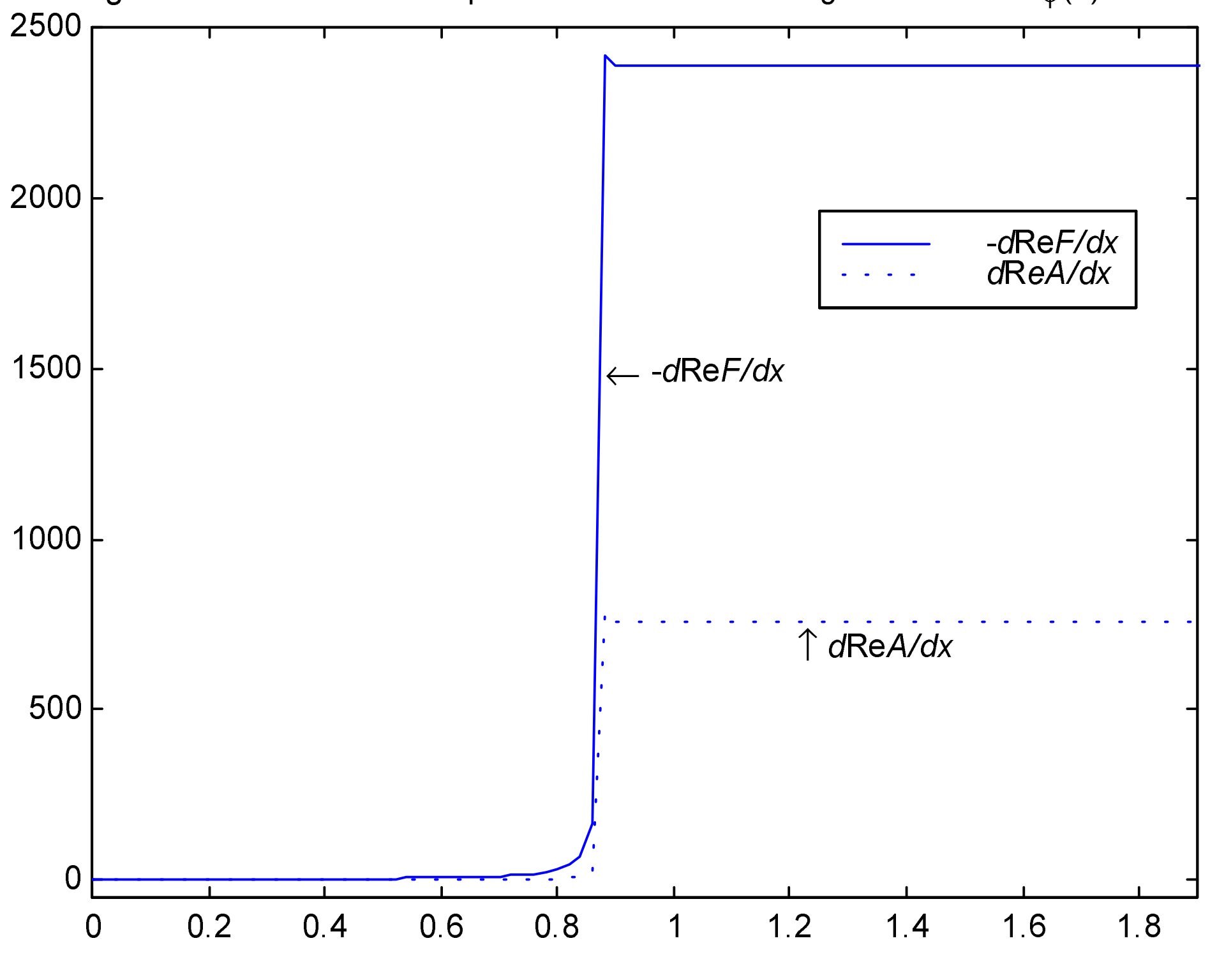

Fig. 14. Profiles of the real parts of electrical and magnetic fields. Im$\psi$(0)=1.0E-2

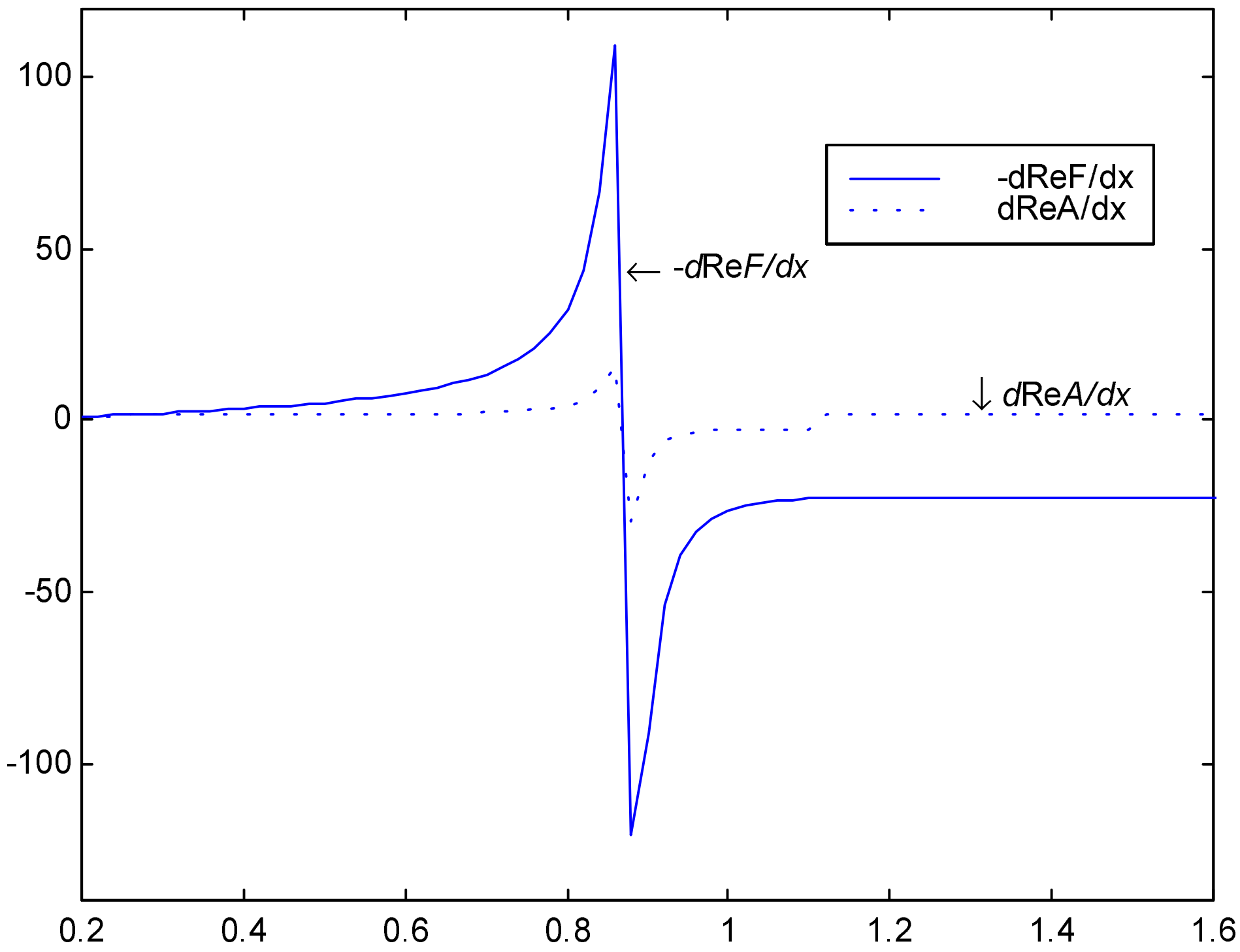

Fig. 15. Profiles of a denseness of cold plasma

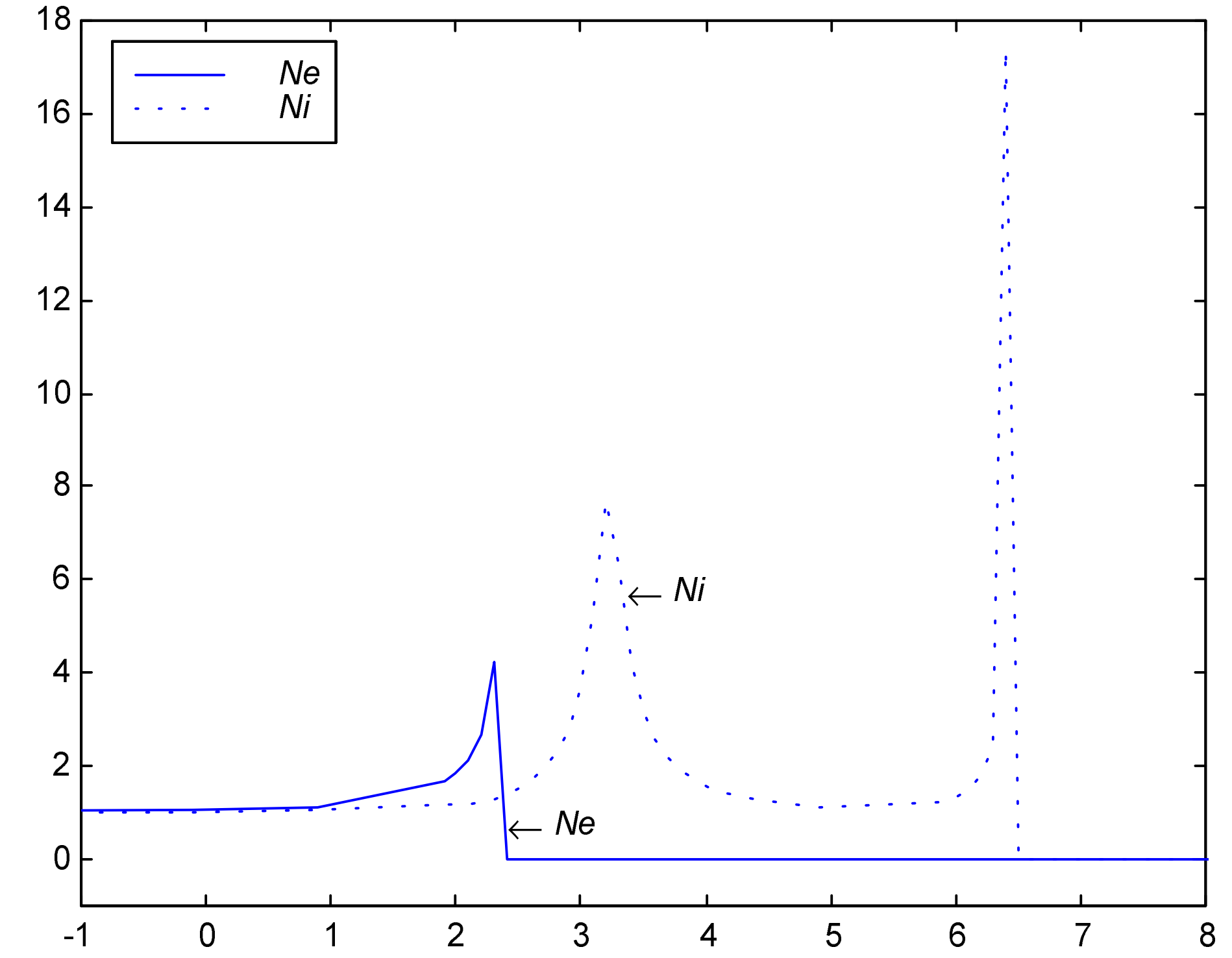

Fig. 16. Profiles of the real part of a denseness of cold plasma Im$\psi$(0)=1.0E-15

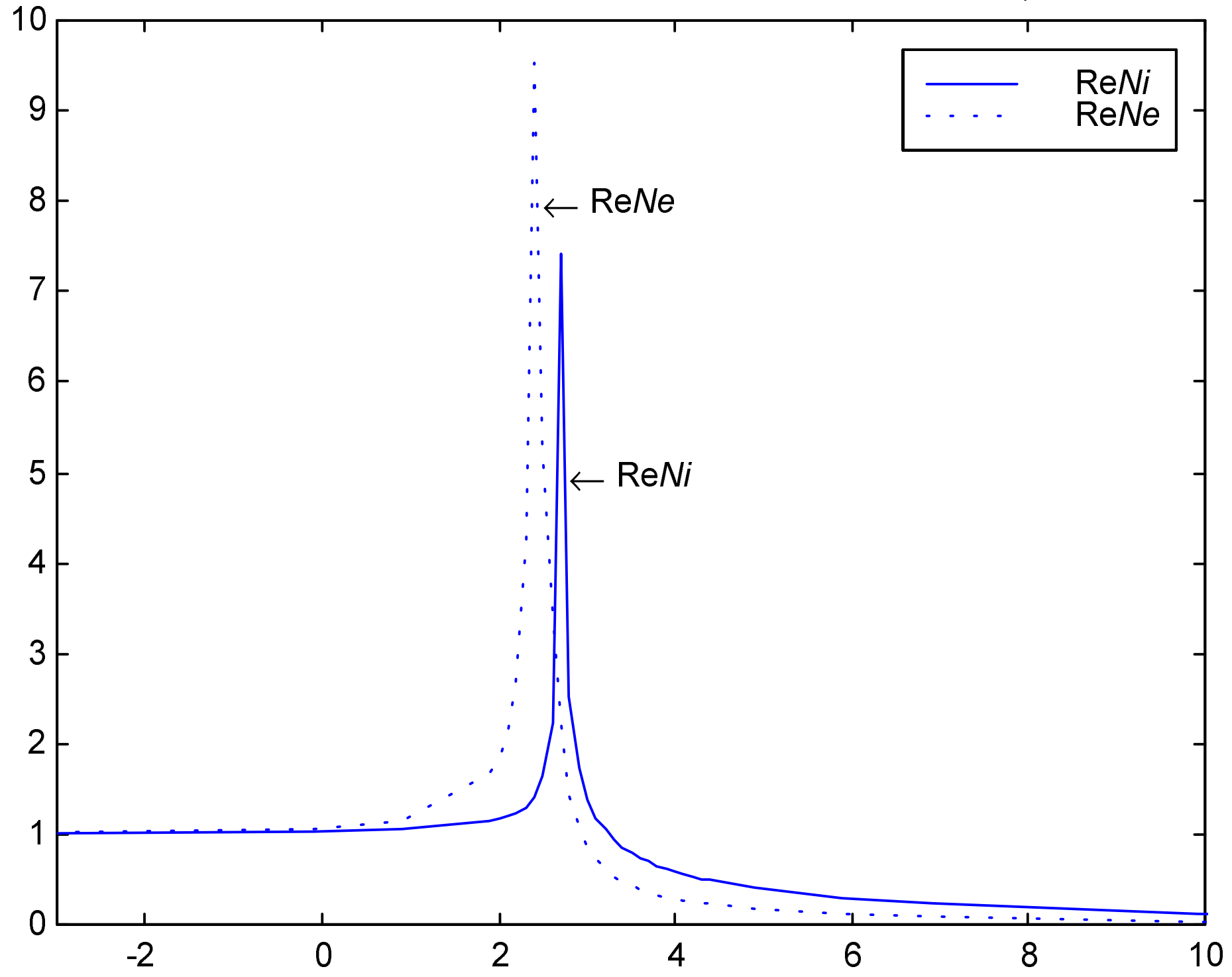